\documentclass[acmsmall,screen,nonacm]{acmart}
\AtBeginDocument{%
  }

\setcopyright{none}
\acmDOI{}
\acmISBN{}

\usepackage{amsthm}
\usepackage[final]{listings}
\usepackage{tikz}
\usetikzlibrary{positioning, arrows.meta, backgrounds}
\usetikzlibrary{shapes, arrows.meta, positioning}
\tikzstyle{block}=[draw,fill=white,rectangle]
\tikzstyle{every node}=[font=\small]
\usepackage{subcaption}
\usepackage{cleveref}
\usepackage{tabularx}
\usetikzlibrary{decorations.pathreplacing}

\lstdefinelanguage{ddlog}{
  language=Java, % we base it on Java, just for comments
  morekeywords={input, output, typedef, relation, typedef, bool, not,
    string, bit, extern, function, var, for, match, skip, in, integer, % not really in DDlog
    Aggregate, FlatMap},
  deletestring=[b]{'}
}

\newcommand{\means}[1]{\ensuremath{\llbracket #1 \rrbracket}}

\newcommand{\N}{\mathbb{N}}  % naturals
\newcommand{\B}{\mathbb{B}}  % Booleans
\newcommand{\stream}[1]{\ensuremath{\mathcal{S}_{#1}}}
\newcommand{\zm}{\ensuremath{z^{-1}}} % stream delay operator
\newcommand{\I}{\mathcal{I}}  % stream integration
\newcommand{\D}{\mathcal{D}}  % stream derivative
\newcommand{\inc}[1]{{#1}^{\Delta}}
\newcommand{\secref}[1]{\S\ref{#1}}  % reference to a section

\newcommand{\lift}[1]{{\uparrow}#1}              
\newcommand{\seq}{\mathbin{\vartriangleright}}
\newcommand{\pll}{\mathbin{||}}

\newcommand{\preserve}{\rightsquigarrow}
\newcommand{\elim}{\smallint}
\usepackage{mathpartir}

\newtheoremstyle{compactdef}
  {0.5\topsep} % 上方间距（可修改为具体数值，如 3pt 或 0pt）
  {0.5\topsep} % 下方间距（可修改为具体数值，如 3pt 或 0pt）
  {\normalfont}% 正文字体（保持为普通罗马体）
  {}           % 整体缩进量（留空表示无缩进）
  {\bfseries}  % 标题字体（设置为加粗 boldface）
  {.}          % 标题后的标点符号
  {0.5em}      % 标题与正文之间的水平间距
  {}           % 标题的额外排版规则（留空则使用默认的 "标题 编号 (可选参数)" 格式）

\theoremstyle{compactdef}

\newtheorem{theorem}{Theorem}[section]

\newtheorem{definition}[theorem]{Definition}
\newtheorem{proposition}[theorem]{Proposition}
\newtheorem{lemma}[theorem]{Lemma}

\makeatletter
\patchcmd{\@thm}{\trivlist}{\trivlist \topsep=0pt \partopsep=0pt}{}{}
\makeatother

\begin{document}

\title{Fixing the Fixpoint:
A Formal Theory of Convergence Detection for Incremental Recursive Computation}

\author{Chengxi Yang}
\authornote{Now at University of California, Berkeley. This work was done
  while an undergraduate at Shanghai Jiao Tong University and a research intern
  at UW-Madison.}
\email{chengxi@berkeley.edu}
\orcid{0009-0009-9302-3351}
\affiliation{%
  \institution{University of Wisconsin--Madison}
  \city{Madison}
  \state{Wisconsin}
  \country{USA}
}
\affiliation{%
  \institution{Shanghai Jiao Tong University}
  \city{Shanghai}
  \country{China}
}

\author{Tej Chajed}
\authornote{Now at Google DeepMind. Work done while at UW-Madison.}
\email{chajed@wisc.edu}
\orcid{0000-0002-9889-4828}
\affiliation{%
  \institution{University of Wisconsin--Madison}
  \city{Madison}
  \state{Wisconsin}
  \country{USA}
}

\author{Thomas Reps}
\email{reps@cs.wisc.edu}
\orcid{0000-0002-5676-9949}
\affiliation{%
  \institution{University of Wisconsin--Madison}
  \city{Madison}
  \state{Wisconsin}
  \country{USA}
}

\renewcommand{\shortauthors}{Yang et al.}

% Current contact institutions, independent of the affiliations for this work.
\authorsaddresses{Authors' Contact Information:
  Chengxi Yang, \nolinkurl{chengxi@berkeley.edu}, University of California, Berkeley;
  Tej Chajed, \nolinkurl{chajed@wisc.edu}, Google DeepMind;
  Thomas Reps, \nolinkurl{reps@cs.wisc.edu}, University of Wisconsin--Madison.}

\begin{abstract}
  Modern incremental computation theories like DBSP have enabled efficient incrementalization of general recursive computations. To do so, they require a runtime Fixpoint Detection (FPD) mechanism to detect whether an iterative computation has reached the fixpoint and thus should terminate.
However, we show that the commonly suggested ``FirstZero'' strategy is unsound even in naturally arising cases, and that exact FPD is impossible for arbitrary DBSP circuits with expressive primitive nodes. This issue reveals a fundamental gap between the mathematical specification and implementations of such theories.
To fill this gap, using DBSP as a core calculus, we develop a formal theory of convergence detection. Within this theory, we define \emph{internal convergence} (IntConv) as a declarative criterion corresponding to the internal-state-stability strategy used by practical implementations, and prove that IntConv is a sufficient condition for external convergence.
We then define the state fixpoint (StFP) predicate and a sound and complete StFP detector. Combining the StFP detector with fixed-input and zero-output checks yields a sound and complete IntConv detector.
Moreover, for a large class of useful circuits (programs) including Datalog queries, nested while queries, and their incrementally optimized versions, we show that IntConv is not only sound but also complete (meaning any convergence in the theory implies the convergence in our criterion). 
As a result, our theory provides semantic guarantees for convergence detection on all these circuits.
Our results are formally verified in Lean, with the formalization available at \url{https://github.com/Arcadia-Y/fixing-the-fixpoint/}.

\end{abstract}




\maketitle

\section{Introduction}\label{sec:intro}
Incremental computation \cite{DBLP:conf/popl/RamalingamR93,DBLP:conf/pepm/Liu24} is a cornerstone of efficient software systems. Its fundamental premise is straightforward: when the input to a computation changes, the system should update the output by processing only the small changes (or ``deltas''), rather than recomputing the entire result from scratch. This principle finds its most classical application in databases in the form of \textit{Incremental View Maintenance} (IVM) \cite{Book:GM99,chirkova2012materialized}, as shown in Fig.~\ref{fig:ivm-and-fixpoint}(a). For standard relational queries, IVM is well-understood and widely deployed, allowing systems to maintain up-to-date views with significantly reduced cost compared to recomputation.

\begin{figure}[!tb]
    \centering
    % Define styles for consistency
    \tikzset{
        main node/.style={font=\large},
        func arrow/.style={->, >={Stealth}, draw=blue!60!cyan, thick},
        update arrow/.style={->, >={Stealth}, draw=red!60!black, thick},
        label text/.style={font=\normalsize}
    }
    \begin{tabular}{c@{\hspace{9.0ex}}c}
        \resizebox{3.0cm}{!}{%
        \begin{tikzpicture}[node distance=1cm and 2cm]
            % Nodes
            \node[main node] (db0) {$DB_0$};
            \node[main node] (v0) [right=of db0] {$V_0$};
            \node[main node] (db1) [below=of db0] {$DB_1$};
            \node[main node] (v1) [right=of db1] {$V_1$};

            % Vertical arrows (Updates) - Continuous lines
            \draw[update arrow] (db0) -- (db1);
            \draw[update arrow] (v0) -- (v1);

            % Top Horizontal arrow (Q)
            \draw[func arrow] (db0) -- node[above, text=black] {$Q$} (v0);

            % Delta Nodes (positioned relative to the center/arrows)
            % ddb is slightly right of the midpoint of the left vertical arrow
            \path (db0) -- coordinate[midway] (midleft) (db1);
            \node (ddb) [right=0.08cm of midleft] {$\Delta DB_1$};

            % dv is slightly left of the midpoint of the right vertical arrow
            \path (v0) -- coordinate[midway] (midright) (v1);
            \node (dv) [left=0.08cm of midright] {$\Delta V_1$};

            % Middle Horizontal arrow (Q^Delta)
            \draw[func arrow] (ddb) -- node[above, text=black] {$Q^\Delta$} (dv);

        \end{tikzpicture}
        }
     &
        \resizebox{7.5cm}{!}{%
        \begin{tikzpicture}[
            node distance=1.4cm,
            main node/.style={font=\large},
            func arrow/.style={->, >={Stealth}, draw=blue!60!cyan, thick},
            label text/.style={font=\small},
            brace style/.style={decoration={brace, mirror, amplitude=10pt, raise=6pt}, decorate, thick}
        ]

            % --- Nodes in a horizontal chain ---
            \node[main node] (db) {$DB$};
            \node[main node] (v1) [right=of db] {$V^1$};
            \node[main node] (v2) [right=of v1] {$V^2$};
            \node[main node] (dots) [right=of v2] {$\dots$};
            \node[main node] (v) [right=of dots] {$V$};

            % --- Arrows with F label ---
            \draw[func arrow] (db) -- node[above] {$F$} (v1);
            \draw[func arrow] (v1) -- node[above] {$F$} (v2);
            \draw[func arrow] (v2) -- node[above] {$F$} (dots);
            \draw[func arrow] (dots) -- node[above] {$F$} node[below, font=\small] {fixpoint} (v);

            % --- Bottom Curly Brace ---
            \draw[brace style] (db.south east) -- (v.south west)
                node[midway, below=16pt] {$Q = \mathsf{fix}\, F$};

        \end{tikzpicture}
        }
    \end{tabular}
    \caption{(a) Incremental view maintenance; (b) iterative fixpoint computation for a recursive query.}\label{fig:ivm-and-fixpoint}
\end{figure}

However, modern data analysis has evolved beyond simple relational queries. Applications ranging from graph analytics to machine learning rely heavily on algorithms that are \textit{iterative or recursive} in nature.
As a simple example,
consider the simple recursive query $Q$ 
for computing the transitive closure of a finite directed graph, expressed in Datalog as follows:

\noindent
{\footnotesize
\begin{lstlisting}[language=ddlog]
// Edge relation with head and tail
input relation E(h: Node, t: Node)
// Reach relation with source s and sink t
output relation R(s: Node, t: Node)
R(x, y) :- E(x, y).
R(x, y) :- E(x, z), R(z, y).
\end{lstlisting}
}

The semantics of this recursive query is defined in terms of the non-recursive query $F_E(R) := E \cup (E \circ R)$.
The result of the recursive query, $Q$, is defined as the
least fixpoint of the equation $x = F_E(x)$.
In this case, the process of finding the least fixpoint can be understood in terms of shortest distances:
if $R^n$ consists of pairs of nodes for which the length of the shortest
connecting path is $\leq n$, then $F_E(R^n) = R^{n+1}$ consists of pairs of nodes whose shortest distance is $\leq n+1$;
consequently, the least fixpoint is the transitive closure of the graph.
This intuition corresponds to the following iterative computation strategy:
repeatedly apply $F_E$
to the accumulated result until a fixpoint is reached, as shown in Fig.~\ref{fig:ivm-and-fixpoint}(b)
(often called \emph{Kleene iteration}).

It is challenging, however, to combine incremental computation with recursion to \emph{efficiently incrementalize recursive queries}---that is, in our example, to update the graph and efficiently recompute the new transitive closure. Luckily, frameworks such as DBSP~\cite{budiu2025dbsp} and Differential Dataflow~\cite{McSherry2013DifferentialD} give elegant theories for efficient incrementalization of general recursive queries.
However, we find that there is a fundamental gap between their mathematical specification and implementations of such frameworks, having to do with detecting fixpoints.
The \emph{Fixpoint Detection} (FPD) problem actually involves two issues:
\begin{enumerate}
  \item
    \label{It:FPDUndecidability}
    If the choice of primitive nodes makes DBSP Turing-complete, exact FPD for arbitrary circuits is \emph{impossible}: there is a fixed level-1 circuit with no computable detector that is both sound and complete (\secref{sec:fpd-undecidability}).
  \item
    \label{It:FirstZeroUnsound}
    In the DBSP paper~\cite{budiu2025dbsp}, the authors suggest that for many practical queries, including Datalog queries, a key primitive for fixpoint detection can be implemented by an iterative procedure that can be stopped when the first ``$0$'' is encountered (e.g., the first $\emptyset$ when working with sets), which intuitively implies that there is no change in two adjacent results.
    We call this the \emph{FirstZero} strategy.
    However, in \secref{sec:overview-problem} (and formalized in
        \secref{sec:fpd-firstzero}), we show that the FirstZero strategy is
        unsound, including in naturally arising cases; that is, using it in the
        implementation results in output that does not match the mathematical specification.
\end{enumerate}

These issues are not confined to DBSP;
as discussed in \secref{sec:related}, similar problems exist in other frameworks supporting incrementalization of general recursive queries, such as Differential Dataflow~\cite{McSherry2013DifferentialD}.

%\ycx{I see, the following points are actually the structure of our argument, but it should not presented as questions to evaluate a solution.}
% Any solution to this problem should be evaluated from the following aspects:
% \begin{itemize}
%   \item
%     Is it \emph{sound}? Will it produce any incorrect results?
%   \item
%     Is it \emph{implementation-aware}? Does it allow efficient implementation?
%   \item
%     For which class of programs is it \emph{complete}? Specifically, will a user ever encounter a scenario where the original theory guarantees convergence, but the implementation fails to do so? 
% \end{itemize}

% In \secref{sec:regular-circuits}, we define the class of \textit{Regular Circuits}, a large class of circuits that DBSP can use to implement general and arbitrarily nested recursive computation.
% Moreover, we show that Regular Circuits are closed under DBSP's incrementalization transformations (\secref{sec:inc-opt}).
% From results given in \secref{sec:detector}, fixpoint detection is decidable for Regular Circuits, and it follows from \secref{sec:complete} that fixpoint detection is efficient for Regular Circuits.

To fill in this gap, we develop a formal theory of convergence detection using DBSP as a core calculus. Within this theory, we define \emph{internal convergence} (IntConv) as a declarative convergence criterion corresponding to the internal-state-stability strategy used by practical implementations such as Feldera~\cite{feldera_repo}.
We show that IntConv is a \emph{sound} criterion, so it never produces an incorrect answer (\secref{sec:intfp}).\footnote{
  Because exact FPD for arbitrary circuits is impossible, IntConv is not complete
  for general circuits.
  Consequently, there can be computations that mathematically converge,
  but convergence is not detected by the IntConv criterion.
}
At the end of \secref{sec:intfp}, we give a theoretical data-dependency example in which relying only on external convergence can require recursive trace-backs and additional retained state. This example motivates an implementation benefit of IntConv.
To connect this declarative criterion to runtime state, we define the State Fixpoint (StFP) predicate and provide a sound and complete StFP detector (\secref{sec:detector}). StFP and the StFP detector give semantic and algorithmic formalizations, respectively, of the internal-state-stability strategy in our denotational model. The global IntConv detector combines the StFP detector with fixed-input and zero-output checks across all bracketed sub-circuits, yielding sound and complete IntConv detection. Because the predicate and StFP detector are structural, our formulation applies compositionally to nested recursive computations; \secref{sec:related} compares this capability with the Feldera implementation~\cite{feldera_repo}.
Moreover, we show that IntConv is
\emph{complete}---i.e., it detects convergence whenever the theoretical specification does---for
a large class of useful circuits
(\secref{sec:useful}). In particular, we use \emph{regular circuits} as a name for the standard DBSP recursive sublanguage generated by lifted scalar circuits, composition, Datalog queries, and while queries. We then show that IntConv is complete for regular circuits and their incrementally optimized versions.\footnote{
  I.e., if a computation of such a circuit mathematically converges, convergence will always be detected by the IntConv criterion.
}
As a result, sound and complete convergence detection is possible for all these useful circuits.

Our key results are formally verified in Lean 4 \cite{Moura2021TheL4}.
Our Lean 4 development is described in \secref{sec:formalization};
all mechanized proofs are available as anonymous supplementary material.

In summary, our contributions are as follows:
\begin{itemize}
  \item
    We identify and formalize the Fixpoint Detection (FPD) problem in DBSP, showing that the commonly used FirstZero criterion is unsound in general and that exact FPD is impossible for arbitrary expressive circuits.
  \item
    We define internal convergence (IntConv) as a declarative criterion corresponding to the internal-state-stability strategy used by Feldera-style implementations, and prove that it is a sound and sufficient condition for convergence.
    We define the State Fixpoint (StFP) predicate and a sound and complete StFP detector, giving semantic and algorithmic formalizations of this strategy. Combining the StFP detector with fixed-input and zero-output checks across all bracketed sub-circuits yields a sound, complete, and compositional global IntConv detector.
  \item
    We show that IntConv is a complete criterion for a large class of useful circuits, including all regular circuits and their incrementally optimized forms. As a result, exact convergence detection is possible for these circuits despite its impossibility for the full language.
  \item
    We formally verify our key results in Lean.
\end{itemize}

\section{Overview}\label{sec:overview}
\subsection{DBSP and the Unsoundness of the FirstZero Strategy}\label{sec:overview-problem}

\begin{figure}[!tb]
    \centering
    \includegraphics[width=\linewidth]{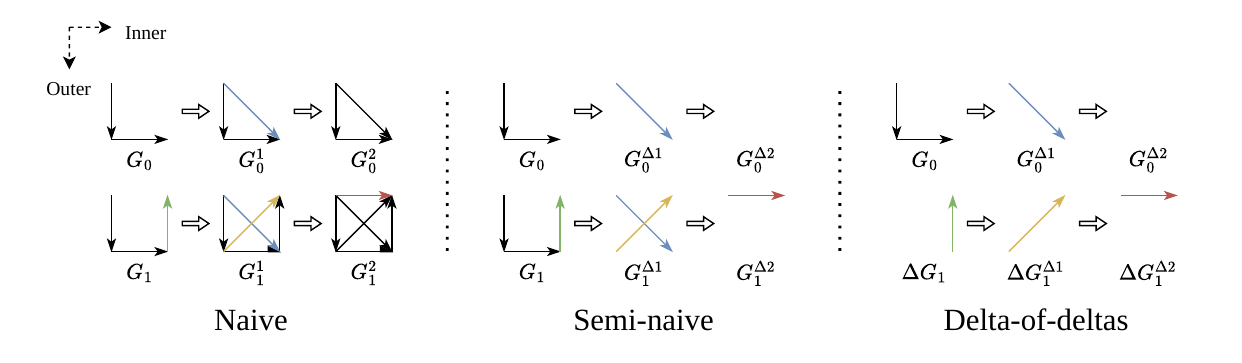}
    \vspace{-4.0ex}
    \caption{Intermediate-evaluation results 
    arising in the three DBSP strategies for incrementalizing a transitive-closure
    query.}\label{fig:ivm-three-steps}
\end{figure}

To lay the groundwork for the problems that we address, we first discuss how DBSP implements and incrementalizes the Datalog query for transitive closure (\secref{sec:intro}) in three steps.
DBSP uses \textit{circuits} to express computations. It first compiles the query into a
``\emph{naive}'' circuit, then optimizes it into a ``\emph{semi-naive}'' circuit, and
finally optimizes it further into a ``\emph{delta-of-deltas}'' circuit. What matters for
our purposes is what is computed incrementally in these circuits;
\Cref{fig:ivm-three-steps} depicts the intermediate results of the evaluation
for one graph and one update to it.
In the figure, the vertical dimension represents changes to the input, i.e., the \emph{outer} iterations for incremental computation (e.g., we add a green edge to the first input $G_0$ to get the second input $G_1$), while the horizontal dimension represents the \emph{inner} iterations for fixpoint computation (e.g., we apply the single-step query $F$ to $G_0$ to get $G_0^1$).
The goal after each outer iteration $m$ is to obtain the transitive-closure
``view'' for graph $G_m$.

The \textbf{naive}
strategy is not optimized for incremental computation at all:
for each outer iteration, it re-runs the entire fixpoint computation from scratch;
on each inner iteration, it merely applies the non-recursive query $F$ to intermediate results (until the inner-iteration fixpoint is obtained).
Its intermediate result in the $n^{\textit{th}}$ inner iteration of the $m^{\textit{th}}$ outer iteration $G_m^n$
is the set of pairs of nodes whose distance is within $n+1$ in $G_m$.

The \textbf{semi-naive} strategy optimizes the naive strategy along the horizontal dimension (inner iterations).
The name
comes from the Datalog semi-naive evaluation strategy (Algorithm 2 from the book~\cite{Greco2015DatalogAL}).
It computes inner-iteration \emph{deltas}
of the intermediate results of the naive strategy (e.g., $G_0^{\Delta 1} = G_0^1 - G_0$ contains only the blue edge).
The intermediate result $G_m^{\Delta n}$, for $n \geq 1$, consists of
pairs of nodes whose distance is exactly $n+1$ in $G_m$.
In each outer iteration, when the inner-fixpoint is reached,
the intermediate results can be collected---e.g., $G_1 \cup G_1^{\Delta 1} \cup G_1^{\Delta 2}$---to obtain the final result.

The \textbf{delta-of-deltas} strategy further incrementalizes the semi-naive strategy
along the outer iterations
by computing \emph{outer-iteration deltas of the inner-iteration deltas}
(e.g., $\Delta G_{1}^{\Delta 1} = G_1^{\Delta 1} - G_0^{\Delta 1}$ contains only the yellow edge).
For each outer iteration, it accepts the delta of the input graph (e.g., $\Delta G_1$), and upon 
obtaining the inner fixpoint, the inner-iteration intermediate results can be collected to obtain the delta of the transitive-closure view.
For instance, for outer iteration 1, the delta of the view is the three pairs in $\Delta G_1 \cup \Delta G_1^{\Delta 1} \cup \Delta G_1^{\Delta 2}$.
Note that this value equals $G_1^2 - G_0^2$.

To produce correct answers, all three strategies
need a common component we call \emph{Fixpoint Detection} (FPD): a runtime mechanism to
detect whether an inner-iteration computation has already reached a fixpoint, at which point, it can terminate the current inner iteration, collect
the intermediate results (for the semi-naive and delta-of-deltas strategies), and proceed to the next outer iteration.

In the DBSP theory,
FPD and inner-iteration collection is encapsulated in
a mathematical operator called \emph{stream elimination}, denoted by $\elim$, which takes an infinite stream $ s \in (\stream{A} := \N \to A)$ representing the intermediate results of an
\emph{unbounded} iterative computation, and produces a scalar value $a \in A$ representing the integrated final result. Intuitively, $\elim(s)$ is defined to return the sum of all \emph{non-zero} elements in $s$.
$\elim$ can only be defined as a \emph{partial} function:
it only produces an answer for
streams that have a \emph{finite} number of non-zero elements (in which case we say
that $\elim(s)$ converges).
When $\elim(s)$ diverges, the implementation can simply run forever or terminate after a timeout.
Either is semantically consistent:
DBSP is designed to be Turing-complete~\cite{budiu2025dbsp} due to unbounded loops.
The question that really matters is ``how can the system detect when a stream $s$
is all zeros beyond some point?'' in which case $\elim(s)$ indeed converges, and the semantics demands that the answer be produced.
The challenge is that the implementation has only a finite prefix of $s$ at run
time, and it must concretely decide all subsequent values of $s$ are zero, but
there are infinitely many of them. Finding a way to establish $s$ has finitely
many non-zero values based on a finite prefix is the essence of
the \textbf{FPD problem} in DBSP.

In the DBSP paper~\cite{budiu2025dbsp}, the authors suggest that for many practical queries including the Datalog queries,  $\elim$ can be ``approximated without loss of precision by integrating until it encounters the first $0$''. We call this 
criterion for deciding convergence the FirstZero criterion (or FirstZero strategy).

For the naive and the semi-naive strategies,\footnote{
  In the DBSP circuit for the naive strategy,
  the intermediate results are in fact first differentiated to become deltas and then sent to $\elim$, so the input to $\elim$ is the same in
  the circuits for the naive and semi-naive strategies.
}
the FirstZero criterion is indeed a sound criterion,
because the first zero in the delta stream ($G_m^{\Delta n} = 0$) implies that the fixpoint of the original stream has been reached ($G_m^{n} = F(G_m^{n-1}) = G_m^{n-1}$), so
$G_m^{\Delta p}$, for $p \geq n$
must all be zero.

\begin{figure}[!tb]
    \centering
    \includegraphics[width=0.46\linewidth]{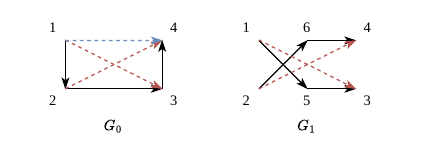}
    \vspace{-4.0ex}
    \caption{
    A pair of input graphs that demonstrate that the FirstZero criterion applied to the delta-of-deltas strategy computes an incorrect answer.
    Instead of the correct transitive closure of $G_1$, namely, $\{ (1,5), (2,6), (5,3), (6,4), \textcolor{red}{(1,3)}, \textcolor{red}{(2,4)} \}$, we would obtain $\{ (1,5), (2,6), (5,3), (6,4), \textcolor{red}{(1,3)}, \textcolor{red}{(2,4)}, \textcolor{blue}{(1,4)} \}$.
    }\label{fig:counterexample}
\end{figure}

However, for the delta-of-deltas strategy, the FirstZero criterion is, in
general, incorrect. That is, if a runtime executes the stream
elimination operator $\elim$ in a DBSP operator by summing until the first zero,
it produces an incorrect result on some inputs where there are more non-zero
elements that should be summed; there is a convergence point in the input, but it is further
ahead in the stream than the first zero.

For a concrete example where FirstZero produces an incorrect answer, consider the two graphs $G_0$ and $G_1$ shown in Fig.~\ref{fig:counterexample}.
In DBSP, the delta between two databases can have negative quantities, representing deletions.
For $G_0$ and $G_1$, the delta $\Delta G_1$ consists of negative edges $\{ (1,2), (2,3), (3,4) \}$ and positive edges $\{ (1,5), (2,6), (5,3), (6,4) \}$.
Note that after the first step of the semi-naive strategy, $G_0^{\Delta 1}$ and $G_1^{\Delta 1}$ are exactly the same (the positive red edges $\{ \textcolor{red}{(1,3)}, \textcolor{red}{(2,4)} \}$).
Consequently, after the first step of the delta-of-deltas strategy, $G_0^{\Delta 1}$ consists of $\{ \textcolor{red}{(1,3)}, \textcolor{red}{(2,4)} \}$ and $\Delta G_{1}^{\Delta 1} = G_1^{\Delta 1} - G_0^{\Delta 1} = \emptyset$ (because subtraction operates over $\mathbb{Z}$-sets, elements with identical positive counts cancel out)---i.e., it is the first zero, so the FirstZero criterion would say, incorrectly, that the inner-iteration fixpoint for $\Delta G_1$ has been reached.

When the delta-of-deltas strategy is properly performed, $\Delta G_1^{\Delta 1}$ is $\emptyset$, but the subsequent $\Delta G_1^{\Delta 2}$ is not $\emptyset$:
$G_0^{\Delta 2}$ consists of the blue edge $\{ \textcolor{blue}{(1,4)} \}$, and thus $\Delta G_1^{\Delta 2}$ consists of the \emph{negative} blue edge $\{ \textcolor{blue}{(1,4)} \}$.
Only when we get to $G_0^{\Delta 3}$ and $\Delta G_1^{\Delta 3}$ do we have $\emptyset$ at corresponding inner-iteration spots of both outer iterations.
The collected inner-iteration intermediate results are
$\Delta G_1 \cup \Delta G_1^{\Delta 1} \cup \Delta G_1^{\Delta 2} \cup \Delta G_1^{\Delta 3}$, which has negative edges $\{ (1,2), (2,3), (3,4), \textcolor{blue}{(1,4)} \}$ and positive edges $\{ (1,5), (2,6), (5,3), (6,4) \}$.
When combined with the transitive closure of $G_0$, i.e., positive edges $\{ (1,2), (2,3), (3,4), \textcolor{red}{(1,3)}, \textcolor{red}{(2,4)}, \textcolor{blue}{(1,4)} \}$, we obtain the transitive closure of $G_1$, $\{ (1,5), (2,6), (5,3), (6,4), \textcolor{red}{(1,3)}, \textcolor{red}{(2,4)} \}$.

In contrast, the FirstZero criterion produces an incorrect result.
The collected inner-iteration intermediate results are
$\Delta G_1 \cup \Delta G_1^{\Delta 1}$, which has negative edges $\{ (1,2), (2,3), (3,4) \}$ and positive edges $\{ (1,5), (2,6), (5,3), (6,4) \}$.
When combined with the transitive closure of $G_0$, i.e., positive edges $\{ (1,2), (2,3), (3,4), \textcolor{red}{(1,3)}, \textcolor{red}{(2,4)}, \textcolor{blue}{(1,4)} \}$, we obtain the incorrect result $\{ (1,5), (2,6), (5,3), (6,4), \textcolor{red}{(1,3)}, \textcolor{red}{(2,4)}, \textcolor{blue}{(1,4)} \}$, which has the extra positive edge $\textcolor{blue}{(1,4)}$.

\subsection{Overview of Our Theory}\label{sec:overview-solution}

The counter example to the soundness of FirstZero as a convergence criterion in \secref{sec:overview-problem}, together with the impossibility result in \secref{sec:fpd-undecidability}, reveals a gap between the mathematical theory of DBSP and its practical implementation. Our theory addresses this gap through a simple progression: specify convergence mathematically, strengthen that specification to require internal stability, and turn internal stability into a runtime check. This progression operates at two scales---locally for one iterative computation and globally for a complete circuit.

To express these two scales uniformly, we define a core calculus of well-formed DBSP circuits (\secref{sec:pre-language}). In this calculus, $\delta_0$ initiates an unbounded iterative computation and $\elim$ integrates its result; we call the circuit between them a \emph{bracketed circuit}. The Fixpoint Detection (FPD) problem asks when one such bracketed circuit reaches a fixpoint. A complete circuit may contain multiple, potentially nested bracketed circuits, so the global \emph{Convergence Detection} problem asks when every bracketed sub-circuit reaches a fixpoint. We formalize both problems and show that exact FPD is impossible for arbitrary circuits with expressive primitive nodes (\secref{sec:fpd}). The result does not apply to Datalog or regular circuits, for which our later completeness theorems yield exact detection.

\begin{figure}[!tb]
    \centering
    \begin{subfigure}[b]{0.58\textwidth}
        \centering
        \includegraphics[width=\linewidth]{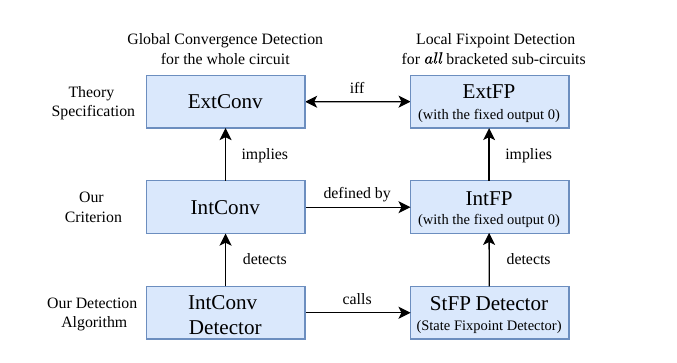}
        \caption{Overview of the IntConv theory.}\label{fig:intconv-overview}
    \end{subfigure}
    \hfill
    \begin{subfigure}[b]{0.40\textwidth}
        \centering
        \includegraphics[width=\linewidth]{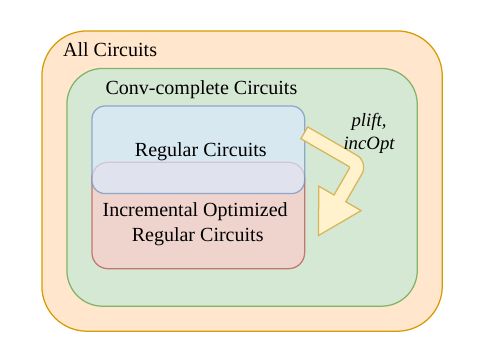}
        \caption{Overview of the Conv-complete circuits.}\label{fig:conv-complete-overview}
    \end{subfigure}
    \caption{Overview of the concepts that play a role in our theory.}\label{fig:overview}
\end{figure}

Fig.~\ref{fig:intconv-overview} organizes the local and global problems into three layers. At the specification layer, \emph{external fixpoint} (ExtFP) says that a circuit's input and output are fixed after a specified iteration; when the fixed output is $0$, ExtFP is equivalent to the original FPD specification. Its global counterpart, \emph{external convergence} (ExtConv), requires every bracketed sub-circuit to converge. At the criterion layer, inspired by the internal-state-stability criterion used in Feldera~\cite{feldera_repo}, \emph{internal fixpoint} (IntFP) strengthens ExtFP by requiring every internal node to reach an external fixpoint, and \emph{internal convergence} (IntConv) requires the inner circuit of every bracketed sub-circuit to satisfy IntFP with fixed output $0$. Consequently, IntFP implies ExtFP and IntConv implies ExtConv (\secref{sec:intfp}). At the detection layer, \emph{state fixpoint} (StFP) connects the internal criterion to the state-oriented check performed at runtime.

The distinction between external and internal convergence also provides theoretical evidence for implementation benefits of IntConv. At the end of \secref{sec:intfp}, we give a data-dependency example in which an evaluation that relies only on ExtConv can require recursive trace-backs and additional retained state. This example is a qualitative example, not an empirical performance evaluation or a general lower bound.

We show that IntConv is indeed detectable using StFP and a high-level detector (\secref{sec:detector}). Although our denotational semantics treats circuits as pure functions, implementations maintain state; StFP asks whether the circuit would be at an internal fixpoint if its observed input remained fixed. We give a structural StFP detector that is sound and complete for StFP. Together with a fixed-input check, it yields a local IntFP detector. The sound and complete global IntConv detector applies this local detector to every bracketed sub-circuit and checks that the fixed output is $0$. Because the StFP detector follows circuit structure, it also applies uniformly to nested bracketed circuits. We will compare this capability with the Feldera implementation in \secref{sec:related}.

Finally, we identify useful classes of circuits on which the internal and external specifications coincide, as shown in Fig.~\ref{fig:conv-complete-overview}. We call a circuit \emph{convergence-complete} (Conv-complete) when its ExtConv implies IntConv for every input. On these circuits, the implementable internal criterion exactly matches the mathematical external convergence specification. We use \emph{regular circuits} as a name for the standard DBSP recursive sublanguage generated by lifted scalar circuits, sequential and parallel composition, Datalog queries, and while queries, including their nested uses. We prove that regular circuits are Conv-complete. We also formalize DBSP's incremental optimization as the syntax-directed transformations $\mathsf{plift}$ and $\mathsf{incOpt}$ and prove that both preserve Conv-completeness, yielding the classes summarized in Fig.~\ref{fig:conv-complete-overview} (\secref{sec:useful}).

\section{Preliminary: A Formal Language for DBSP}\label{sec:preliminary}
% Background section: DBSP stream computation fundamentals
In this section, we first introduce the core concepts of DBSP that are necessary to understand the FPD problem and our solution (\secref{sec:pre-theory}), then present our formal language for DBSP well-formed circuits (\secref{sec:pre-language}), and finally show how the Datalog query from \secref{sec:intro} is implemented and incrementalized as an example (\secref{sec:pre-example}).

\subsection{Streams and Operators in DBSP}\label{sec:pre-theory}
In this section, we introduce the core DBSP theory, which forms the universe for our language's semantics.

\begin{definition}[Stream]\label{def:stream}
Given an Abelian group $(A, +, 0_A, -)$, a \emph{stream} of values from $A$, or an $A$-stream, is a function $s : \mathbb{N} \to A$. We denote the set of all $A$-streams by $\mathcal{S}_A = \{ s \mid s : \mathbb{N} \to A \}$. We write $s[t]$ for the $t$-th element of stream $s$.
\end{definition}
Streams inherit the group structure pointwise: $(s_1 + s_2)[t] = s_1[t] + s_2[t]$, so $(\mathcal{S}_A, +, 0_{\mathcal{S}_A}, -)$ is itself an Abelian group.

\begin{definition}[Stream Operator]\label{def:stream-operator}
A \emph{stream operator} is a function $T : \mathcal{S}_{A_1} \times \cdots \times \mathcal{S}_{A_n} \to \mathcal{S}_B$.
\end{definition}

In circuit diagrams, we depict stream operators as blocks and streams as arrows.

\begin{definition}[Lifting]\label{def:lifting}
Given a function $f : A \to B$, its \emph{lifting} ${\uparrow} f : \mathcal{S}_A \to \mathcal{S}_B$ applies $f$ pointwise in time: $({\uparrow} f)(s)[t] = f(s[t])$. Lifting distributes over composition: ${\uparrow}(f \circ g) = ({\uparrow} f) \circ ({\uparrow} g)$.
\end{definition}

\begin{definition}[Delay]
The \emph{delay} operator $z^{-1} : \mathcal{S}_A \to \mathcal{S}_A$ shifts a stream by one time step, inserting zero at $t = 0$: $z^{-1}(s)[t] = 0_A$ if $t = 0$, and $s[t-1]$ for $t \geq 1$.
\end{definition}

\begin{definition}[Feedback Cycle]\label{def:feedback-cycle}
For certain operators $T: (\stream{A} \times \stream{B}) \to \stream{B}$ and $F: \stream{B} \to \stream{B}$, the equation $Q(s) = T(s, F(Q(s)))$ has a unique solution $Q: \stream{A} \to \stream{B}$, denoted as $\lambda s.\, \mathrm{fix}\, \alpha.\, T(s, F(\alpha))$.\footnote{Our formal language guarantees that all feedback cycles are well-defined.} Feedback cycles can be represented as Fig.~\ref{fig:feedback-cycle}.
\end{definition}

DBSP uses feedback cycles mainly for two purposes: express iterative computation and construct integration operator $\I$, as we will show in the following.

\begin{figure}[ht]
    \centering
    \begin{subfigure}[b]{0.3\textwidth}
        \centering
        \begin{tikzpicture}[>=latex, node distance=1cm]
            \node[] (input) {$s$};
            \node[block, right of=input] (f) {$T$};
            \node[right of=f] (output) {$\alpha$};
            \node[block, below of=f, node distance=.5cm] (z) {$F$};
            \draw[->] (input) -- (f);
            \draw[->] (f) -- node (mid) {} (output);
            \draw[->] (mid.center) |-  (z);
            \draw[->] (z.west) -- ++(-.3,0) |- ([yshift=1mm]f.south west);
        \end{tikzpicture}
        \caption{Feedback cycle}
        \label{fig:feedback-cycle}
    \end{subfigure}
    \hfill
    \begin{subfigure}[b]{0.3\textwidth}
        \centering
        \begin{tikzpicture}[auto,>=latex]
            \node[] (input) {$s$};
            \node[block, shape=circle, right of=input, inner sep=0pt] (plus) {$+$};
            \node[right of=plus, node distance=1.5cm] (output) {$\I(s)$};
            \node[block, below of=plus, node distance=.7cm] (z) {$z^{-1}$};
            \draw[draw,->] (input) -- (plus);
            \draw[->] (plus) -- node (o) {} (output);
            \draw[->] (o) |- (z);
            \draw[->] (z) -- (plus);
        \end{tikzpicture}   
        \caption{Integration $\I$}
        \label{fig:integration}
    \end{subfigure}
    \hfill
    \begin{subfigure}[b]{0.3\textwidth}
        \centering
        \begin{tikzpicture}[auto,>=latex,node distance=1cm]
            \node[] (input) {$s$};
            \node[block, shape=circle, right of=input, inner sep=0pt,node distance=2cm] (plus) {$+$};
            \node[right of=plus] (output) {$\D(s)$};
            \draw[draw,->] (input) -- node (i) {} (plus);
            \node[block, below of=i, node distance=.7cm] (z) {$\zm$};
            \node[block, shape=circle, right of=z, inner sep=1pt] (minus) {$-$};
            \draw[->] (plus) -- (output);
            \draw[->] (i) -- (z);
            \draw[->] (z) -- (minus);
            \draw[->] (minus) -- (plus);
        \end{tikzpicture}
        \caption{Differentiation $\D$}
        \label{fig:differentiation}
    \end{subfigure}
    \caption{Some core DBSP operators represented as circuit diagrams.}
    \label{fig:core-circuits}
\end{figure}

\begin{definition}[Integration and Differentiation]\label{def:integration}
The \emph{integration} operator $\I : \mathcal{S}_A \to \mathcal{S}_A$ is $\I(s) = \mathrm{fix}\, \alpha.\, (s + z^{-1}(\alpha))$, computing the running sum: $\I(s)[t] = \sum_{i \leq t} s[i]$, as shown in Fig.~\ref{fig:integration}.
The \emph{differentiation} operator $\D : \mathcal{S}_A \to \mathcal{S}_A$ is $\D(s) = s - z^{-1}(s)$, as shown in Fig.~\ref{fig:differentiation}.
\end{definition}

$\I$ and $\D$ are inverses ($\I \circ \D = \D \circ \I = \mathrm{id}$).
Given a stream operator $Q : \mathcal{S}_A \to \mathcal{S}_B$, its \emph{incremental version} is defined as $Q^\Delta = \D \circ Q \circ \I$.
The operator $Q^\Delta$ consumes a stream of changes and produces a stream of changes: if $Q(s_1) = s_2$, then $\inc{Q}(\D(s_1))= \D(s_2)$. 
The DBSP theory establishes compositional properties for $Q^\Delta$, such as the chain rule $(Q_1 \circ Q_2)^\Delta = Q_1^\Delta \circ Q_2^\Delta$. This leads to a syntax-guided incrementalization algorithm, which we will formalize as $\mathsf{incOpt}$ in \secref{sec:pre-language}.

To incrementalize iterative computations (like recursive Datalog queries), we need to introduce a new ``inner'' time dimension for iterative fixpoint computation besides the ``outer'' time dimension for incremental computation, as \secref{sec:overview-problem} shows. DBSP implements this via two specific operators and nested streams.

\begin{definition}[Stream Introduction]\label{def:stream-intro}
The \emph{stream introduction} operator $\delta_0 : A \to \mathcal{S}_A$ produces a stream from a scalar value $v$:
$\delta_0(v)[0] = v$ and $\delta_0(v)[t] = 0$ for $t > 0$.
\end{definition}

\begin{definition}[Stream Elimination]\label{def:stream-elim}
The \emph{stream elimination} operator $\textstyle\elim : \mathcal{S}_A \to A$ aggregates a stream that is \emph{zero almost everywhere} (i.e., $\exists t_0 \in \mathbb{N}.\, \forall t \geq t_0.\, s[t] = 0_A$) into a scalar value: $\textstyle\elim(s) = \sum_{t \geq 0} s[t]$.
\end{definition}
As we mentioned in \secref{sec:overview-solution}, $\delta_0$ takes a scalar input and outputs a stream and initiates a new unbounded iterative computation. $\elim$ integrates the intermediate results of the inner computation back into a single scalar output.
If the input stream of $\elim$ is not zero almost everywhere, the summation diverges. This potential for divergence leads to the FPD problem that we address in this paper.

A \emph{nested stream} (or stream of streams) has type $\mathcal{S}_{\mathcal{S}_A} = \mathbb{N} \to (\mathbb{N} \to A)$. It can be viewed as an infinite 2D matrix where $(t_0, t_1)$ index the outer time and inner time, respectively, and each row is an inner stream. 

We call an operator $S : \mathcal{S}_{\mathcal{S}_A} \to \mathcal{S}_{\mathcal{S}_B}$ a \emph{nested operator}. Just as we lift scalar functions to operators, we can lift operators to nested operators. If $S : \mathcal{S}_A \to \mathcal{S}_B$, then ${\uparrow}S : \mathcal{S}_{\mathcal{S}_A} \to \mathcal{S}_{\mathcal{S}_B}$ applies $S$ to each \emph{row} of the nested stream: $({\uparrow}S)(s)[t_0] = S(s[t_0])$. For a scalar function $f:A \to B$, we lift it twice to get $\lift{\lift{f}}: \stream{\stream{A}} \to \stream{\stream{B}}$, defined as $(\lift{\lift{f}})(s)[m][n] = f(s[m][n])$.

For nested streams, delay $z^{-1}$ shifts values along the outer time dimension, delaying ``rows'' of the matrix, while lifted delay ${\uparrow}z^{-1}$ shifts values along the \emph{inner} time dimension, delaying ``columns'' of the matrix. The integration $\I$ operates along the outer time dimension, summing up rows of the matrix, while the lifted integration ${\uparrow}\I$ operates along the inner time dimension, summing up columns of the matrix.

\subsection{A Formal Language for DBSP Well-Formed Circuits}\label{sec:pre-language}

We now formalize a language of DBSP \emph{well-formed circuits} (which we simply abbreviate as \emph{circuits} from now on) that serves as the core calculus for our theory.
This circuit language is an inductive grammar as shown in Table~\ref{tab:circuit-syntax}.

A circuit $c : A \to B$ transforms an input stream of type $A$ to an output stream of type $B$. Types are built from \emph{base types} (Abelian groups) and binary products $A \times B$.
Each circuit is annotated with a \emph{stream level} $\ell \in \{1, 2\}$,\footnote{\label{fn:stream-level}
  This restriction to the stream level is mainly for formalization convenience without loss of expressiveness. See \secref{sec:formalization} for more discussion on this point.
} representing whether the circuit's input and output are streams (level 1) or nested streams (level 2). For instance, a $\mathsf{bracket}$ node must wrap a level-2 circuit and output a level-1 circuit.

The \emph{primitive node} construct ($\mathsf{node}$) lifts a user-defined scalar function to a stream operator, providing a generic extension point for core computations. Commonly used generic operations (like identity, tuple projections, and group arithmetic) are separated into a class of \emph{standard nodes}.

\begin{table}[htb]
    \centering
    \caption{Circuit syntax and denotational semantics. Constructs are polymorphic over stream level $\ell$ unless indicated.}
    \label{tab:circuit-syntax}
    \footnotesize
    \begin{tabular}{llllll}
        \toprule
        \textbf{Construct $c$} & \textbf{Type} & \textbf{Level} & \textbf{$\llbracket c \rrbracket$} & \textbf{$\mathsf{plift}(c)$} & \textbf{$\mathsf{incOpt}(c)$} \\
        \midrule
        \multicolumn{6}{l}{\textit{Primitive nodes}} \\
        $\mathsf{node}(f)$ & $A \to B$ & $\ell$ & ${\uparrow}^{\ell} f$ & $\mathsf{node}(f)$ & $\mathsf{incOpt}(\mathsf{node}(f))$ \\
        \midrule
        \multicolumn{6}{l}{\textit{Standard nodes}} \\
        $\mathsf{const}(x)$ & $A \to B$ & $\ell$ & ${\uparrow}^{\ell} (\lambda\_.\ x)$ & $\mathsf{const}(x)$ & $\mathsf{const}(x) \seq \mathsf{D}$ \\
        $\mathsf{id}$ & $A \to A$ & $\ell$ & $\mathrm{id}$ & $\mathsf{id}$ & $\mathsf{id}$ \\
        $\mathsf{fst}$ & $A \times B \to A$ & $\ell$ & ${\uparrow}^{\ell} \pi_1$ & $\mathsf{fst}$ & $\mathsf{fst}$ \\
        $\mathsf{snd}$ & $A \times B \to B$ & $\ell$ & ${\uparrow}^{\ell} \pi_2$ & $\mathsf{snd}$ & $\mathsf{snd}$ \\
        $\mathsf{add}$ & $A \times A \to A$ & $\ell$ & ${\uparrow}^{\ell} (+)$ & $\mathsf{add}$ & $\mathsf{add}$ \\
        $\mathsf{sub}$ & $A \times A \to A$ & $\ell$ & ${\uparrow}^{\ell} (-)$ & $\mathsf{sub}$ & $\mathsf{sub}$ \\
        \midrule
        \multicolumn{6}{l}{\textit{Temporal nodes}} \\
        $\mathsf{delay}$ & $A \to A$ & $\ell$ & $z^{-1}$ & $\mathsf{\lift{delay}}$ & $\mathsf{delay}$ \\
        $\mathsf{\lift{delay}}$ & $A \to A$ & $2$ & $\lift{z^{-1}}$ & $-$ & $\mathsf{\lift{delay}}$ \\
        \midrule
        \multicolumn{6}{l}{\textit{Structural combinators}} \\
        $c_1 \seq c_2$ & $A \to C$ & $\ell$ & $\llbracket c_2 \rrbracket \circ \llbracket c_1 \rrbracket$ & $\mathsf{plift}(c_1) \seq \mathsf{plift}(c_2)$ & $\mathsf{incOpt}(c_1) \seq \mathsf{incOpt}(c_2)$ \\
        $c_1 \pll c_2$ & $A \to B \times C$ & $\ell$ & $\lambda x.\, (\llbracket c_1 \rrbracket(x),\, \llbracket c_2 \rrbracket(x))$ & $\mathsf{plift}(c_1) \pll \mathsf{plift}(c_2)$ & $\mathsf{incOpt}(c_1) \pll \mathsf{incOpt}(c_2)$ \\
        \midrule
        \multicolumn{6}{l}{\textit{Feedback loops}} \\
        $\mathsf{loop}(c)$ & $A \to B$ & $\ell$ & $\lambda x.\, \mathrm{fix}\, \alpha.\, \means{c}(x, z^{-1}(\alpha))$ & $\mathsf{\lift{loop}}(\mathsf{plift}(c))$ & $\mathsf{loop}(\mathsf{incOpt}(c))$ \\
        $\mathsf{\lift{loop}}(c)$ & $A \to B$ & 2 & $\lambda x.\, \mathrm{fix}\, \alpha.\, \means{c}(x, \lift{z^{-1}}(\alpha))$ & $-$ & $\mathsf{\lift{loop}}(\mathsf{incOpt}(c))$ \\
        \midrule
        \multicolumn{6}{l}{\textit{Level adapters}} \\
        $\mathsf{lifting}(c)$ & $A \to B$  & 2 $(c \text{ at } 1)$ & ${\uparrow}\llbracket c \rrbracket$ & $-$ & $\Delta(\mathsf{lifting}(c))$ \\
        $\mathsf{bracket}(c)$ & $A \to B$  & 1 $(c \text{ at } 2)$ & $\lift{\elim} \circ \means{c} \circ \lift{\delta_0}$ & $\mathsf{lifting}(\mathsf{bracket}(c))$ & $\mathsf{bracket}(\mathsf{incOpt}(c))$ \\
        \bottomrule
    \end{tabular}
\end{table}

Several important circuits can be defined compositionally: 
\begin{definition}[Derived Circuit Constructs]
    The integration circuit $\mathsf{I} = \mathsf{loop}(\mathsf{add})$ computes $\I$, and the differentiation circuit $\mathsf{D} = (\mathsf{id} \pll \mathsf{delay}) \seq \mathsf{sub}$ computes $\D$. The unoptimized incremental form of a circuit $c$ is given by $\Delta(c) = \mathsf{I} \seq c \seq \mathsf{D}$.
\end{definition}

The denotational semantics $\llbracket c \rrbracket$ of a circuit $c : A \to B$ at stream level $\ell$ is a stream operator. For level 1, it is an operator $\mathcal{S}_A \to \mathcal{S}_B$. For level 2, it is $\mathcal{S}_{\mathcal{S}_A} \to \mathcal{S}_{\mathcal{S}_B}$. Here ${\uparrow}^{1} f = {\uparrow} f$ denotes single lifting, and ${\uparrow}^{2} f = {\uparrow}{\uparrow} f$ denotes double lifting.

Due to the $\mathsf{bracket}$ construct, which relies on the stream elimination operator $\elim$, the denotational semantics $\llbracket \cdot \rrbracket$ is a \emph{partial} function. Whether it converges on a given input is the convergence detection problem we mentioned in \secref{sec:overview-solution}.

We can now formalize the incremental optimization algorithm of DBSP as the following two syntactic functions:

\begin{definition}\label{def:plift}
    $\mathsf{plift}$ lifts a level-1 circuit and pushes the lifting operator onto individual nodes. Its definition is in Table~\ref{tab:circuit-syntax}.
\end{definition}
\begin{definition}\label{def:incOpt}
    $\mathsf{incOpt}$ transforms a circuit into its optimized incremental version following the DBSP's syntax-guided algorithm. Its definition is in Table~\ref{tab:circuit-syntax}.
\end{definition}

Note that for primitive nodes, $\mathsf{incOpt}(\mathsf{node}(f))$ is a customizable incremental form, which makes this algorithm quite extensible. The theory has provided an incremental form for a wide class of functions. For example, for a linear function $f$, the incremental form can be $\mathsf{node}(f)$ itself. Even in the worst case, the incremental form can fall back to the unoptimized form $\Delta(\mathsf{node}(f))$.

Another thing to notice is that $\mathsf{incOpt}(\mathsf{lifting}(c))$ is the unoptimized form, as the incremental operator $\Delta$ cannot go across $\mathsf{lifting}$ into the inner circuit. That's why we need the function $\mathsf{plift}$ to push the lifting operator into the inner circuit.

\subsection{Example: Implementing and Incrementalizing the Datalog Query in DBSP}\label{sec:pre-example}

In this section, we present how DBSP expresses and incrementalizes the Datalog query example from \secref{sec:intro} and \secref{sec:overview-problem} in three steps (corresponding to the three strategies in \secref{sec:overview-problem}).

DBSP uses \textbf{$\mathbb{Z}$-sets} \cite{Green2009ReconcilableD} (or simply Zsets) as the underlying Abelian group to represent sets and relations. It can be simply understood as sets where each element has an integer count, which can be negative to represent deletion. In this paper, we will simply treat Zsets as their underlying sets for most cases.

\begin{figure}[ht]
    \centering
    \begin{subfigure}[b]{0.35\textwidth}
        \centering
        \begin{tikzpicture}[>=latex, node distance=.72cm]
            \node[] (Iinput) {$i$};
            \node[block, right of=Iinput] (Idelta) {$\delta_0$};
            \node[block, right of=Idelta] (I) {$\I$};
            \node[block, right of=I] (f) {$\lift{F}$};
            \node[block, right of=f, node distance=1.0cm] (D) {$\D$};
            \node[block, right of=D] (S) {$\elim$};
            \node[right of=S] (output)  {$o$};
            \draw[->] (f) -- node (o) {} (D);
            \node[block, below of=o, node distance=.55cm] (z) {$\zm$};
            \draw[->] (Iinput) -- (Idelta);
            \draw[->] (S) -- (output);
            \draw[->] (o.center) -- (z);
            \draw[->] (z) -| (f);
            \draw[->] (Idelta) -- (I);
            \draw[->] (I) -- (f);
            \draw[->] (D) -- (S);
        \end{tikzpicture}
        \caption{The naive strategy.}
        \label{fig:naive-dbsp}
    \end{subfigure}
    \hfill
    \begin{subfigure}[b]{0.28\textwidth}
        \centering
        \begin{tikzpicture}[>=latex, node distance=.97cm]
            \node[] (Iinput) {$i$};
            \node[block, right of=Iinput, node distance=0.7cm] (Idelta) {$\delta_0$};
            \node[block, right of=Idelta] (f) {$\inc{(\lift{F})}$};
            \node[block, right of=f, node distance=1.1cm] (S) {$\elim$};
            \node[right of=S, node distance=.7cm] (output)  {$o$};
            \draw[->] (f) -- node (o) {} (S);
            \node[block, below of=o, node distance=.65cm] (z) {$\zm$};
            \draw[->] (Iinput) -- (Idelta);
            \draw[->] (S) -- (output);
            \draw[->] (o.center) -- (z);
            \draw[->] (z) -| (f);
            \draw[->] (Idelta) -- (f);
        \end{tikzpicture}   
        \caption{The semi-naive strategy.}
        \label{fig:semi-naive-dbsp}
    \end{subfigure}
    \hfill
    \begin{subfigure}[b]{0.35\textwidth}
        \centering
        \begin{tikzpicture}[>=latex,node distance=.95cm]
            \node[] (Iinput) {$\vec{i}$};
            \node[block, right of=Iinput, node distance=.75cm] (Idelta) {$\lift{\delta_0}$};
            \node[block, right of=Idelta, node distance=1.35cm] (f) {$\inc{(\lift{\inc{(\lift{F})}})}$};
            \node[block, right of=f, node distance=1.5cm] (S) {$\lift{\elim}$};
            \node[right of=S, node distance=.75cm] (output)  {$\vec{o}$};
            \draw[->] (f) -- node (o) {} (S);
            \node[block, below of=o, node distance=.65cm] (z) {$\lift{\zm}$};
            \draw[->] (Iinput) -- (Idelta);
            \draw[->] (S) -- (output);
            \draw[->] (o.center) -- (z);
            \draw[->] (z) -| (f);
            \draw[->] (Idelta) -- (f);
        \end{tikzpicture}
        \caption{The delta-of-deltas strategy.}
        \label{fig:delta-of-deltas-dbsp}
    \end{subfigure}
    \caption{The three circuits that arise when DBSP implements and incrementalizes a Datalog recursive query.}
    \label{fig:three-approaches-DBSP}
\end{figure}

Recall that the recursive query is defined via a non-recursive query $F_E(R)$, or $F(E, R)$ (which we will simply write as $F$). DBSP will first compile it into some circuit and lift it to get a level-1 circuit $c_0$  whose denotation is $\lift{F}$.
The recursive query $Q$ asks for the fixpoint of iterative application of $F$ on an input $i$. $Q$ can be expressed as a circuit shown in Fig.~\ref{fig:naive-dbsp}. 
The query body (the bracketed sub-circuit) can be expressed in our formal language as\footnote{
    Technically, we cannot express the whole level-0 circuit itself in our formal language because of our syntactic restriction (see \secref{sec:formalization}), but it suffices to express the query body as it encodes sufficient information.
}
\begin{equation}\label{eq:naive-body}
    \mathsf{datalogBody}(c_0) = \mathsf{loop}(\Delta(c_0))
\end{equation}
We use the stream introduction operator $\delta_0$ (Def.~\ref{def:stream-intro}) and the integration operator $\I$ (Def.~\ref{def:integration}) to convert the input $i$ into a constant stream of $i$ and then feed it into $\lift{F}$. The feedback loop implements the iterative fixpoint computation we need. Its output stream $\alpha$ looks like following:
\[ \alpha[0] = F(i, 0), \ \alpha[1] = F(i, \alpha[0]), \ \alpha[2] = F(i, \alpha[1]), \ \ldots \] 
If $\alpha$ is fixed after some point, then the differentiation operator $\D$ (Def.~\ref{def:integration}) converts it into a stream that is zero almost everywhere, which can then be aggregated by the stream elimination operator $\elim$ (Def.~\ref{def:stream-elim}) to produce the final output $o$. This corresponds to the naive strategy in \secref{sec:overview-problem}.

Note that we can optimize $\lift{F}$ between $\I$ and $\D$ into $\inc{(\lift{F})}$ by applying the DBSP's incremental optimization algorithm, naturally leading to the circuit implementing the semi-naive strategy shown in Fig.~\ref{fig:semi-naive-dbsp}. Its query body can be expressed as
\begin{equation}\label{eq:semi-naive-body}
    \mathsf{semiNaiveBody}(c_0) = \mathsf{loop}(\mathsf{incOpt}(c_0))
\end{equation}

We can further lift the whole circuit, push the lifting operator and apply the incremental optimization to transform it into the circuit shown in Fig.~\ref{fig:delta-of-deltas-dbsp}, which implements the delta-of-deltas strategy.
Its query body can be expressed as
\begin{equation}\label{eq:delta-of-deltas-body}
    \mathsf{deltaOfDeltasBody}(c_0) = \mathsf{\lift{loop}}\Bigl(\mathsf{incOpt} \Bigl(\mathsf{plift} \bigl(\mathsf{incOpt}(c_0)\bigr) \Bigr) \Bigr)
\end{equation}
and the whole circuit can be expressed as
\begin{equation}\label{eq:opt-datalog-query}
    \mathsf{deltaOfDeltasQuery}(c_0) = \mathsf{bracket}(\mathsf{deltaOfDeltasBody}(c_0))
\end{equation}

\section{The Fixpoint-Detection Problem and the Convergence-Detection Problem}\label{sec:fpd}
As introduced in \secref{sec:pre-theory}, the stream-elimination operator $\elim$ is inherently a partial function: it is well-defined only if its input stream is \emph{zero almost everywhere}.
With standard Zset-based relational operators, DBSP is Turing-complete \cite{budiu2025dbsp}, meaning it is acceptable that some circuits containing $\elim$ never converge to zero.
However, the core challenge is at run time: even if $\elim$'s input does eventually converge to zero, how can the system safely detect this convergence by observing only a finite prefix of the stream? Determining this zero-almost-everywhere property at run time is the essence of the \emph{Fixpoint Detection (FPD)} problem. In this section, we formalize the FPD problem and the \emph{soundness} and \emph{completeness} of fixpoint detectors.

Furthermore, in our language (\secref{sec:pre-language}), $\elim$ operators are packaged inside bracketed circuits, and a complete circuit can contain multiple bracketed sub-circuits, much like a standard program can contain multiple while loops. Therefore, a circuit mathematically \emph{converges} if and only if every bracketed sub-circuit reaches a fixpoint. We refer to this global detection problem as the \emph{Convergence Detection} problem.

To formalize the FPD problem within our formal language, we introduce the following definition to
characterize when
a stream has converged to zero at a specific index:
\begin{definition}[ZeroAfter]\label{De:ZeroAfter}
    A stream $s \in \stream{A}$ is \emph{zero after an outer index $n \in \N$} if:
    \[\text{ZeroAfter}_1(s, n) := \forall m \ge n, s[m] = 0_A\]
    A nested stream $s \in \stream{\stream{A}}$ is \emph{zero after a 2D-index $(m, n) \in \N \times \N$} if:
    \[\text{ZeroAfter}_2(s, m, n) := \text{ZeroAfter}_1(s[m], n)\]
    We also say that a nested stream $s \in \stream{\stream{A}}$ is \emph{zero after an upper-bound stream $\mathbf{b} \in \stream{\N}$} if:
    \[\text{ZeroAfter}_2(s, \mathbf{b}) := \forall i \in \N, \text{ZeroAfter}_2(s, i, \mathbf{b}[i])\]
\end{definition}

An important fact is that ZeroAfter is a \emph{level-indexed predicate} with the following properties:
\begin{itemize}
    \item $\text{ZeroAfter}_1$ applies to any stream (i.e., stream of level $\ell \ge 1$), and accepts a 1D-index.
    \item $\text{ZeroAfter}_2$ is defined in terms of $\text{ZeroAfter}_1$. It applies only to nested streams (i.e., streams of level $\ell \ge 2$), and accepts a 2D-index or a stream of 1D-indices.
\end{itemize}
This inductive\footnote{
    One might consider Def.~\ref{De:ZeroAfter} to not be inductive because it only refers to levels 1 and 2.
    However, this choice was deliberate because of the syntactic restriction we placed on stream level (see \secref{sec:pre-language} and \secref{sec:formalization}). If the stream level is generalized to any natural number, Def.~\ref{De:ZeroAfter} would be a standard inductive definition.
} structure is a natural result of the stream-level hierarchy of a well-defined circuit.     We will soon see more examples of such level-indexed definitions. 

We also need to introduce another predicate to capture the first true value of a stream of Booleans.
\begin{definition}[FirstTrue]
    A stream of Booleans $s \in \stream{\B}$ encounters its first true value at index $n \in \N$, denoted $\text{FirstTrue}_1(s) = n$, if $s[n]= \text{true}$ and $\forall m < n, s[m] = \text{false}$. Similarly, we define $\text{FirstTrue}_2(s) = \mathbf b$ for $s \in \stream{\stream{\B}}$ and $\mathbf b \in \stream{\N}$ as $\forall i \in \N, \text{FirstTrue}_1(s[i]) = \mathbf b[i]$.
\end{definition}

We can now formally state the FPD problem. Intuitively, the FPD problem is a matter of providing a fixpoint detector that specifies a runtime fixpoint-detection mechanism for a level-1 bracketed circuit $c$ (as shown in Fig.~\ref{fig:fpd-1}) or a level-2 bracketed circuit $c$ (as shown in Fig.~\ref{fig:fpd-2}). A fixpoint detector accepts a circuit $c$ and an input to this circuit (which is either $\delta_0(x)$ or $\lift{\delta_0}(x)$), and returns a Boolean value for each entry of the input. When a fixpoint detector returns true for the first time in an iteration, the output of $c$ should have converged to $0$ in the corresponding iteration (recall that $c$ should produce changes to the output that are integrated by $\elim$ until convergence). We call this property \textit{soundness}
(see Eqns.~\eqref{eq:fpd-sound} and \eqref{eq:fpd-sound-2} below).
When the output has converged to $0$, the fixpoint detector should eventually return true. We call this property \textit{completeness}
(see Eqns.~\eqref{eq:fpd-complete} and \eqref{eq:fpd-complete-2} below).
\begin{definition}[The FPD problem]\label{def:fpd-problem}
    Given a level-1 circuit $c$, a function $f_c: \stream{A} \to \stream{\B}$ is a \textit{fixpoint detector} for $c$ if for any $x \in A$ and $n \in \N$:
    \begin{equation}\label{eq:fpd-sound}
        \text{FirstTrue}_1(f_c(\delta_0(x))) = n \implies \text{ZeroAfter}_1(\means{c}(\delta_0(x)), n) \text{ and}
    \end{equation}
    \begin{equation}\label{eq:fpd-complete}
        \text{ZeroAfter}_1(\means{c}(\delta_0(x)), n) \implies \exists m, \text{FirstTrue}_1(f_c(\delta_0(x))) = m
    \end{equation}
    Similarly, given a level-2 circuit $c$, a function $f_c: \stream{\stream{A}} \to \stream{\stream{\B}}$ is a \textit{fixpoint detector} for $c$ if
    for any $x \in \stream{A}$ and $\mathbf{b} \in \stream{\N}$:
    \begin{equation}\label{eq:fpd-sound-2}
        \text{FirstTrue}_2(f_c(\lift{\delta_0}(x))) = \mathbf{b} \implies \text{ZeroAfter}_2(\means{c}(\lift{\delta_0}(x)), \mathbf{b}) \text{ and}
    \end{equation}
    \begin{equation}\label{eq:fpd-complete-2}
        \text{ZeroAfter}_2(\means{c}(\lift{\delta_0}(x)), \mathbf{b}) \implies \exists \mathbf{b}', \text{FirstTrue}_2(f_c(\lift{\delta_0}(x))) = \mathbf{b}'
    \end{equation}
\end{definition}

\begin{figure}[t]
    \centering
    \begin{subfigure}[b]{0.45\textwidth}
        \centering
        \begin{tikzpicture}[>=latex]
            \node[] (Iinput) {$x$};
            \node[block, right of=Iinput] (Idelta) {$\delta_0$};
            \node[block, right of=Idelta] (c) {$c$};
            \node[block, right of=c] (S) {$\elim$};
            \node[right of=S] (output)  {};
            \draw[->] (c) -- node (o) {} (S);
            \draw[->] (Iinput) -- (Idelta);
            \draw[->] (Idelta) -- (c);
            \draw[->] (c) -- (S);
            \draw[->] (S) -- (output);
        \end{tikzpicture}
        \caption{The FPD problem for a level-1 circuit $c$}
        \label{fig:fpd-1}
    \end{subfigure}
    \hfill
    \begin{subfigure}[b]{0.45\textwidth}
        \centering
        \begin{tikzpicture}[>=latex]
            \node[] (Iinput) {$x$};
            \node[block, right of=Iinput] (Idelta) {$\lift{\delta_0}$};
            \node[block, right of=Idelta] (c) {$c$};
            \node[block, right of=c] (S) {$\lift{\elim}$};
            \node[right of=S] (output)  {};
            \draw[->] (c) -- node (o) {} (S);
            \draw[->] (Iinput) -- (Idelta);
            \draw[->] (Idelta) -- (c);
            \draw[->] (c) -- (S);
            \draw[->] (S) -- (output);
        \end{tikzpicture}
        \caption{The FPD problem for a level-2 circuit $c$}
        \label{fig:fpd-2}
    \end{subfigure}
    \caption{The FPD problem for level-1 and level-2 circuits.}
    \label{fig:fpd-ckt}
\end{figure}
By changing FirstZero to
another
fixpoint specification or criterion (like our IntFP to be introduced in \secref{sec:intfp}), Def.~\ref{def:fpd-problem} can be adapted to formalize other FPD problems.
Note that the FPD for a level-1 circuit $c$ on some input $x$ can be simply reduced to the FPD for the level-2 circuit $\lift{c}$ on stream input $\lambda\_.\ x$.
Consequently,
level-2 FPD is a harder problem than level-1 FPD.

For the theory's global convergence specification, we define it as follows:
\begin{definition}[External convergence (ExtConv)]
    For any circuit $c$ and input $x$, we say $c$ \emph{externally converges}, denoted $\text{ExtConv}(c, x)$, if $\means{c}(x)$ mathematically converges. Or equivalently, for any sub-circuit in the form of $\mathsf{bracket}(c_0)$ and its corresponding input $x_0$, $\exists \mathbf{b},\, \text{ZeroAfter}_2(\means{c_0}(\lift{\delta_0}(x_0)), \mathbf{b})$.
\end{definition}

It is easy to see why the external-convergence problem is just the global version of the FPD problem. Once we have a fixpoint detector for any circuit $c$, we immediately get a convergence detector for ExtConv by applying the fixpoint detector to every bracketed sub-circuit. Therefore, we will only focus on the FPD problem in
what follows.

\subsection{Example: The FirstZero Strategy and Its Limitation}\label{sec:fpd-firstzero}

We have already introduced the FirstZero strategy in Section~\ref{sec:overview-problem}, which is
merely to stop
when the first $0$ value is encountered. 
We can formalize FirstZero by instantiating the $f_c$ fixpoint-detector function 
from
Def.~\ref{def:fpd-problem}. For a level-1 circuit $c$, it maps an input stream $x$ to a Boolean stream: $f_c(x) = \lambda n.\, (\means{c}(x)[n] = 0)$. For a level-2 circuit $c$, it maps an input nested stream $x$ to a nested Boolean stream: $f_c(x) = \lambda i.\, \lambda n.\, (\means{c}(x)[i][n] = 0)$.

Following the intuitive idea in \secref{sec:overview-problem}, we have established the following theorems for 
the soundness and completeness of FirstZero for the naive and semi-naive strategy, and the unsoundness of FirstZero for the delta-of-deltas strategy.

\begin{theorem}\label{thm:firstzero-valid}
    The FirstZero strategy is a sound and complete fixpoint detector for the naive query body $\mathsf{datalogBody}(c_0)$ (Eqn.~(\ref{eq:naive-body})) and the semi-naive query body $\mathsf{semiNaiveBody}(c_0)$ (Eqn.~(\ref{eq:semi-naive-body})), where $c_0$ can represent any circuit implementing a lifted scalar query.
\end{theorem}

\begin{theorem}\label{thm:firstzero-counterexample}
    The FirstZero strategy is not a sound fixpoint detector for the delta-of-deltas query body $\mathsf{deltaOfDeltasBody}(c_0)$ (Equation~\ref{eq:delta-of-deltas-body}), where $c_0$ is the circuit implementing the lifted non-recursive query $F_E(R)$ from \secref{sec:intro}.
\end{theorem}

\subsection{Impossibility of Exact Fixpoint Detection}\label{sec:fpd-undecidability}

Can another strategy overcome FirstZero's failure and solve FPD exactly for every circuit? The following theorem shows the answer is no. We assume the choice of primitive nodes makes DBSP \emph{Turing-complete}: DBSP circuits can express arbitrary partial-recursive scalar functions. This assumption is mild in practice: standard Zset primitive nodes together with DBSP's recursive constructs suffice~\cite[\S7.2]{budiu2025dbsp}. Without loss of generality, we also assume the base type is \(\mathbb Z\).

\begin{theorem}\label{thm:fpd-undecidability}
    If the choice of primitive nodes makes DBSP Turing-complete, there exists a fixed level-1 circuit $c$ with no computable sound and complete fixpoint detector.
\end{theorem}
\begin{proof}
    We show that a computable sound and complete detector would recognize\footnote{In computability theory, an algorithm recognizes a set if the algorithm eventually accepts every input in the set and never accepts inputs outside it; on those other inputs, it may run forever.} the complement of the halting problem, which is not recursively enumerable. We construct a fixed circuit whose output is eventually zero exactly for inputs that encode nonhalting programs. Let $h(e,k)$ be $1$ if the program that $e$ encodes halts on input $0$ within $k$ steps of a fixed universal evaluator, and $0$ otherwise. Since $h$ is total and primitive recursive~\cite[\S5.2]{carneiro2019computability}, Turing-completeness ensures there is a level-1 circuit $E$ with $\means{E}=\lift{h}$. Define
    \(
        c := \bigl(\mathsf{I} \pll (\mathsf{const}(1) \seq \mathsf{I})\bigr) \seq E.
    \)
    On $\delta_0(e)$, the two branches supply $e$ and $n+1$, giving $\means{c}(\delta_0(e))[n]=h(e,n+1)$. Assume there is a computable sound and complete detector for $c$ and run the detector on \(\delta_0(e)\) until its first true output. By completeness, this procedure terminates if \(e\) does not halt; by soundness, it runs forever if \(e\) halts. This would recognize the complement of the halting problem, a contradiction.
\end{proof}

The theorem only rules out exact fixpoint detection for \emph{arbitrary circuits}, but it does not rule out sound and complete detectors for \emph{certain circuit subclasses of interest}.\footnote{For example, this construction does not establish an impossibility result for incrementalizations of total computable queries, because the query \(\elim(\llbracket c\rrbracket(\delta_0(e)))\) is defined exactly for nonhalting programs.} It motivates the sound sufficient criterion IntConv developed in the next section. In \secref{sec:useful}, we prove that IntConv is also complete for regular circuits, including Datalog and while queries, and then extend this guarantee to their incrementally optimized forms.

\section{A Sound Convergence Criterion: IntConv}\label{sec:intfp}
In this section, we introduce \emph{internal convergence}
(IntConv)
as a sound sufficient condition for ExtConv.
(In \secref{sec:detector}, we give an implementable detector for the IntConv criterion.)
Following the ``local-global'' structure of the problem, we will first define \emph{internal fixpoint} (IntFP) as a sound sufficient condition for an external fixpoint, and then
apply it
to all bracketed sub-circuits to 
formulate
the IntConv criterion.
At the end of this section, we give a qualitative data-dependency example in which relying on ExtConv without IntConv requires recursive trace-backs or additional retained state. The example provides a theoretical argument for an implementation benefit of IntConv.

Before introducing IntFP, we first introduce a natural generalization of \text{ZeroAfter} called \text{FixAfter}, which indicates that a stream becomes fixed to a constant value rather than to $0_A$.
\begin{definition}[FixAfter]
    Like \text{ZeroAfter}, \text{FixAfter} is a level-indexed predicate defined identically in its level-1, level-2, and vectorized forms, except that its base case $\text{FixAfter}_1(s, n)$ requires $\forall m \ge n,\, s[m] = s[n]$. Note that if $s$ is a nested stream $\stream{\stream{A}}$, this means repeating the exact same \emph{inner stream} $s[n]$ from the outer index $n$ onwards.
\end{definition}

Then, we reframe the FPD problem using a new level-indexed predicate called external fixpoint (ExtFP). Intuitively, ExtFP characterizes that a circuit reaches an external fixpoint at a given index, without looking into the internal circuit structure. Formally, 

\begin{definition}[External fixpoint (ExtFP)]
    Given any circuit $c$ (either level-1 or level-2), an input stream $x$, and an index $n \in \N$, we say that $c$ reaches an \emph{external fixpoint} across the first time dimension at $n$, denoted by $\text{ExtFP}_1(c, x, n)$, if both the input and the denoted output of $c$ are fixed after the outer index $n$:
    \[ \text{ExtFP}_1(c, x, n) := \text{FixAfter}_1(x, n) \land \text{FixAfter}_1(\means{c}(x), n) \]

    For a level-2 circuit $c$, a nested input stream $x \in \stream{\stream{A}}$, an outer index $m \in \N$, and an inner index $n \in \N$, the external fixpoint across the second time dimension is defined as:
    \[ \text{ExtFP}_2(c, x, m, n) := \text{FixAfter}_2(x, m, n) \land \text{FixAfter}_2(\means{c}(x), m, n) \]
    $\text{ExtFP}_2(c, x, \mathbf{b})$ for $\mathbf{b} \in \stream{\N}$ is naturally defined as $\forall i \in \N, \text{ExtFP}_2(c, x, i, \mathbf{b}[i])$.
\end{definition}

With this definition, the FPD problem for a level-1 circuit $c$ on input $x$ essentially asks to detect an index $n$ where $\text{ExtFP}_1(c, \delta_0(x), n)$ holds and the fixed output is $0$. Similarly, the FPD problem for a level-2 circuit $c$ on nested input $x$ asks to detect a bound stream $\mathbf{b}$ where $\text{ExtFP}_2(c, \lift{\delta_0}(x), \mathbf{b})$ holds and the fixed outputs are $0$.

As we proved in \secref{sec:fpd-undecidability}, exact FPD is impossible for sufficiently expressive circuits, and hence so is detecting ExtFP in full generality.
The Feldera implementation of DBSP~\cite{feldera_repo} uses internal-state stability to detect fixpoints. Each runtime operator provides a scope-indexed fixed-point check intended to report whether its output will remain stable if its input does, and the circuit combines the checks of its operators. Thus, the operational idea of detecting internal-state stability is established implementation practice rather than a novel contribution of this paper. The cited implementation specifies and tests this runtime contract, but does not provide a formal proof that the circuit-wide check implies denotational convergence.
To bridge the gap between this runtime strategy and the denotational DBSP theory, we give the strategy a rigorous, declarative account through a criterion called \emph{internal fixpoint} (IntFP). This formal criterion allows us to prove precisely this soundness connection, and later establish completeness for a large class of circuits. We discuss the relationship with the Feldera implementation further in \secref{sec:related}.

\begin{definition}[Internal fixpoint (IntFP)]\label{De:IntFP}
    Intuitively, IntFP means that all internal nodes (i.e., individual components) of a circuit have reached an ExtFP at a specific index.
    Formally, for any circuit $c$, an input stream $x$, and an index $n \in \N$, $\text{IntFP}_1(c, x, n)$ is defined by structural induction on $c$ over the categories in Table~\ref{tab:circuit-syntax}:
    \begin{itemize}
        \item \textbf{Primitive, Standard and Temporal Nodes:} Inherit the $\text{ExtFP}_1$ condition. That is, for any such node $c$, $\text{IntFP}_1(c, x, n) := \text{ExtFP}_1(c, x, n)$.
        \item \textbf{Structural Combinators:} Require $\text{IntFP}_1$ for all sub-circuits. That is
        \[\text{IntFP}_1(c_1 \seq c_2, x, n) := \text{IntFP}_1(c_1, x, n) \land \text{IntFP}_1(c_2, \means{c_1}(x), n)\] 
        and 
        \[\text{IntFP}_1(c_1 \pll c_2, x, n) := \text{IntFP}_1(c_1, x, n) \land \text{IntFP}_1(c_2, x, n)\].
        \item \textbf{Feedback Loops:} Require $\text{IntFP}_1$ for the loop body on the input tuple containing the delayed output. That is
        \[\text{IntFP}_1(\mathsf{loop}(c), x, n) := \text{IntFP}_1(c, (x, z^{-1}(\means{\mathsf{loop}(c)}(x))), n)\] 
        and similarly 
        \[\text{IntFP}_1(\mathsf{\lift{loop}}(c), x, n) := \text{IntFP}_1(c, (x, \lift{z^{-1}}(\means{\mathsf{\lift{loop}}(c)}(x))), n)\]
        \item \textbf{Level Adapters:} For $\mathsf{bracket}(c)$, $\text{IntFP}_1$ delegates to the level-2 internal circuit $c$: \[\text{IntFP}_1(\mathsf{bracket}(c), x, n) := \text{IntFP}_1(c, \lift{\delta_0}(x), n)\]
        For the level-1 circuit $\mathsf{lifting}(c)$, because it models an inner circuit that lacks states across outer iterations, the internal fixpoint naturally degenerates into the external fixpoint: 
        \[\text{IntFP}_1(\mathsf{lifting}(c), x, n) := \text{ExtFP}_1(\mathsf{lifting}(c), x, n)\]
    \end{itemize}
    The predicate $\text{IntFP}_2(c, x, m, n)$ for the second time dimension is defined for level-2 circuits only. It is mostly defined by uniformly replacing $\text{ExtFP}_1$ with $\text{ExtFP}_2$ and $\text{IntFP}_1$ with $\text{IntFP}_2$, with a notable exception. For the nested lifting circuit $\mathsf{lifting}(c)$, IntFP relies on the first dimension predicate $\text{IntFP}_1$ of the underlying level-1 circuit: 
    \[\text{IntFP}_2(\mathsf{lifting}(c), x, m, n) := \text{IntFP}_1(c, x[m], n)\]
    The vectorized version is naturally defined as $\text{IntFP}_2(c, x, \mathbf{b}) := \forall i \in \N, \text{IntFP}_2(c, x, i, \mathbf{b}[i])$.
\end{definition}

The definition of IntFP is both structural and level-indexed. For a bracketed circuit, it applies IntFP to the enclosed circuit, adapting the input to its stream level. Because the same rules apply recursively to sub-circuits, the definition also handles nested brackets.

One of the most important properties of IntFP is that it is a sound sufficient condition for $\text{ExtFP}$. That is, if a circuit reaches an internal fixpoint, it must have also reached an external fixpoint.

\begin{theorem}[Soundness of IntFP]\label{thm:intfp-sound}
    For any well-formed circuit $c$, input $x$, and index $n$, we have:
    \[ \text{IntFP}_1(c, x, n) \implies \text{ExtFP}_1(c, x, n) \]
    Similarly, for any level-2 circuit $c$, nested input $x$, and indices $m, n$:
    \[ \text{IntFP}_2(c, x, m, n) \implies \text{ExtFP}_2(c, x, m, n) \]
\end{theorem}
\begin{proof}
    We first prove the property for the level index 1, and then prove the property for the level 2 based on that for level 1. This can also be seen as induction on the level index of the predicate.
\end{proof}

Finally, by applying the local IntFP criterion to all bracketed sub-circuits, and adding that the fixed output is 0, we define our global convergence criterion called \emph{internal convergence} (IntConv).

\begin{definition}[Internal convergence (IntConv)]\label{def:intconv}
    We say that a circuit $c$ internally converges on input $x$, denoted $\text{IntConv}(c, x)$, if for any sub-circuit in the form of $\mathsf{bracket}(c_0)$ and its input $x_0$,
    \[\exists \mathbf{b},\, \text{IntFP}_2(c_0, \lift{\delta_0}(x_0), \mathbf{b}) \land \forall i, \means{c_0}(\lift{\delta_0}(x_0))[i][\mathbf{b}[i]] = 0\]
\end{definition}

By Theorem~\ref{thm:intfp-sound}, the definition of IntConv will be equivalent with $\forall i, \means{c_0}(\lift{\delta_0}(x_0))[i][\mathbf{b}[i]] = 0$ replaced by $\text{ZeroAfter}_2(\means{c_0}(\lift{\delta_0}(x_0)), \mathbf{b})$. This immediately implies that IntConv is a sound sufficient condition for ExtConv.
\begin{theorem}[Soundness of IntConv]\label{thm:intconv-sound}
    For any well-formed circuit $c$ and input $x$, we have
    \[\text{IntConv}(c, x) \implies \text{ExtConv}(c, x)\]
\end{theorem}

These soundness implications are one-way in general: ExtFP need not imply IntFP, and ExtConv need not imply IntConv. In \secref{sec:useful}, we identify circuit classes for which the corresponding converse does hold.

Why use IntConv rather than a weaker sound criterion for the theoretical specification ExtConv? The following example isolates one implementation benefit. For this circuit, an evaluator that relies only on ExtConv lacks a guarantee that earlier internal values have stabilized. It must therefore either recompute a recursive dependency chain or retain the historical values needed to answer later demands. IntConv supplies the missing stability guarantee.

\begin{figure}[!tb]
    \centering
    \begin{subfigure}[b]{0.48\textwidth}
        \centering
        \includegraphics[width=\linewidth]{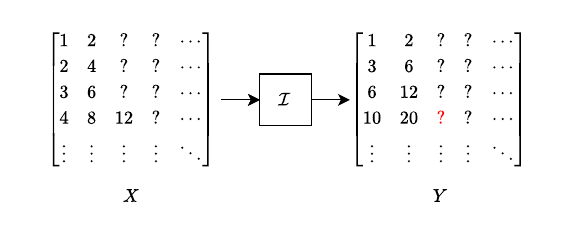}
        \caption{An $\I$ operator inside a bracketed sub-circuit. The red \textcolor{red}{?} is the value we want to evaluate. }\label{fig:intconv-efficient-1}
    \end{subfigure}
    \hfill
    \begin{subfigure}[b]{0.48\textwidth}
        \centering
        \includegraphics[width=\linewidth]{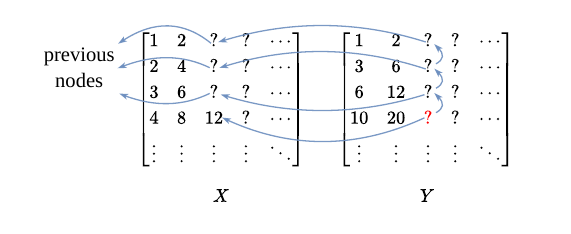}
        \caption{Data dependency in this example. An arrow from $a$ to $b$ denotes that computing $a$ requires $b$.}\label{fig:intconv-efficient-2}
    \end{subfigure}
    \caption{A data-dependency cost that can arise when evaluation relies on ExtConv without IntConv.}\label{fig:intconv-efficient}
\end{figure}

Consider such a circuit satisfying ExtConv but not IntConv, with an integration operator $\mathcal{I}$ inside a bracketed sub-circuit, as shown in Fig.~\ref{fig:intconv-efficient}.
Its input $X$ and output $Y$ are both nested streams.
Assume ExtConv is detected soundly.
Suppose that evaluation is currently at $(4,3)$ (we use 1-based indexing)---namely, the third inner iteration of the fourth outer iteration---and that the first three outer iterations reached external convergence after two inner iterations, while the fourth did not.
Now $X[4][3]=12$ is ready.

Fig.~\ref{fig:intconv-efficient-2} shows the data dependency in this example. To evaluate $Y[4][3]$ correctly, we need
$Y[4][3] = Y[3][3] + X[4][3]$.
But because the circuit satisfies ExtConv without satisfying IntConv, external convergence does not guarantee internal convergence.
Consequently, we cannot directly conclude what $Y[3][3]$ is.
Computing $Y[3][3]$ may require $Y[2][3]$ and $X[3][3]$, both unknown; computing $X[3][3]$ can recursively force upstream re-evaluation, potentially all the way back to $\lift{\delta_0}$,
which initiated
the iterative computation.

Thus, under a strategy that does not retain all of this history, a later outer iteration that needs more inner iterations than earlier ones can trigger recursive recomputation. Retaining the history instead trades that recomputation for additional state, potentially across an unbounded number of outer iterations.

For this example, IntConv avoids that tradeoff.
Because IntConv guarantees that all internal nodes reach an internal fixpoint, values like $Y[3][3]$ are mathematically guaranteed to be stabilized values (e.g., $Y[3][3]=Y[3][2]$ in this example), which should already be stored in the state.
No recursive trace-back is needed, and state management stays simple: in this example, $\mathcal I$ only needs to store its output list (all output until convergence) from the last outer iteration.
This analysis is qualitative: it neither establishes a general lower bound for every criterion weaker than IntConv nor measures the overhead of a concrete implementation.

\section{The StFP Detector and the Global IntConv Detector}\label{sec:detector}
According to Def.~\ref{def:intconv}, the global IntConv detector checks every bracketed sub-circuit with a local IntFP detector and also checks that its fixed output is $0$. We construct the local detector in two steps: the StFP detector checks local state stability, and a fixed-input check turns its result into IntFP detection. Thus, the global IntConv detector combines the StFP detector with fixed-input and zero-output checks across all bracketed sub-circuits.

To detect the IntFP, the Feldera implementation for the DBSP runtime uses a scope-indexed internal-state-stability strategy that we call ``\emph{FirstStateStable}'', as opposed to FirstZero. The idea in our core calculus is as follows: most nodes are stateless, and only $z^{-1}$ and $\lift{z^{-1}}$ are stateful. For stateless nodes, their output is solely determined by their input. For stateful $z^{-1}$ and $\lift{z^{-1}}$, their output in an execution step is solely determined by their state and unrelated to the input of the current step, which means we can predict all stateful nodes' outputs at the end of the previous step. Therefore, at the end of a step, if we can tell that the outputs of all stateful nodes of a circuit are the same as the previous step, and that the input to the whole circuit is fixed after this step
(which is easily decidable for input like $\delta_0(x)$), then we can conclude that the circuit has reached an IntFP.

To formalize this idea, a challenge is that for simplicity, our theory only specifies the denotational semantics of DBSP circuits as pure mathematical functions, and has no formal concepts of ``states''. After all, although we have some model in mind of how states are implemented, ultimately it is a representation choice in the implementation.
Luckily, there is an important assumption that holds for any reasonable implementation of states: \emph{a circuit's state at a given point is determined by its input up to that point}. Therefore, we can always formalize properties of a circuit's state in terms of its input up to that point.

Following this idea, we first make precise how to continue an observed input by holding it fixed. For a stream $x$ and an index $n$, let
\[
\mathsf{letFixed}_1(x,n)[i] :=
\begin{cases}
    x[i] & i < n,\\
    x[n] & i \geq n,
\end{cases}
\]
and, for a nested stream $x$ and indices $m,n$, let
\[
\mathsf{letFixed}_2(x,m,n)[i][j] :=
\begin{cases}
    x[m][n] & i=m \land j \geq n,\\
    x[i][j] & \text{otherwise}.
\end{cases}
\]
Thus, $\mathsf{letFixed}_2$ leaves every row other than $m$ unchanged and replaces the suffix of row $m$ beginning at $n$ with the value $x[m][n]$.

We can now define the notion of \emph{State Fixpoint (StFP)} as a semantic counterpart of the circuit-state property used by the FirstStateStable strategy.
\begin{definition}[State Fixpoint (StFP)]
    For any circuit $c$, input $x$, and index $n$, define
    \[
    \text{StFP}_1(c,x,n) :=
    \text{IntFP}_1(c,\mathsf{letFixed}_1(x,n),n).
    \]
    For any level-2 circuit $c$, nested input $x$, and indices $m,n$, define
    \[
    \text{StFP}_2(c,x,m,n) :=
    \text{IntFP}_2(c,\mathsf{letFixed}_2(x,m,n),m,n).
    \]
    Finally, for a bound stream $\mathbf{b} \in \stream{\N}$, the vectorized predicate is defined pointwise:
    \[
    \text{StFP}_2(c,x,\mathbf{b}) :=
    \forall i \in \N,\, \text{StFP}_2(c,x,i,\mathbf{b}[i]).
    \]
\end{definition}

The fixed continuations agree with the observed input through the designated index and impose a fixed future from that index onward. Therefore, StFP asks whether the circuit's current state would be at an internal fixpoint if its future input remained stable. The following theorem states its precise connection to IntFP.
\begin{theorem}[IntFP and StFP]
    For any circuit $c$,
    \[\forall x, n, \, \text{IntFP}_1(c, x, n) \iff (\text{StFP}_1(c, x, n) \land \text{FixAfter}_1(x, n))\]
    and for any level-2 circuit $c$,
    \[\forall x, m, n, \, \text{IntFP}_2(c, x, m, n) \iff (\text{StFP}_2(c, x, m, n) \land \text{FixAfter}_2(x, m, n))\]
    The pointwise definitions immediately give the vectorized form
    \[\forall x, \mathbf{b}, \, \text{IntFP}_2(c, x, \mathbf{b}) \iff (\text{StFP}_2(c, x, \mathbf{b}) \land \text{FixAfter}_2(x, \mathbf{b})).\]
\end{theorem}
Because it is easy to detect whether $\lift{\delta_0}(x)$ (the input to the target bracketed circuit) is fixed after any index $(m, n)$, it suffices to detect the StFP.

We now give a high-level runtime algorithm to detect the StFP for \emph{any} circuit $c$ and input $x$. While StFP provides the semantic formalization of FirstStateStable in our denotational model, this detector provides its algorithmic formalization. As before, we simulate computation of the state at some point by computation of input up to that point. We will also explain what a natural implementation would probably do for each case. See \secref{sec:related} for more discussion of the Feldera implementation.

\begin{definition}[StFP Detector]
    We define $\text{Detector}_1$, the level-1 StFP detector that accepts a circuit $c$, stream input $x$, and an index $n$, and returns a boolean, as follows:
    \begin{itemize}
        \item \textbf{Primitive and Standard Nodes:} Return true because they are stateless. 
        \[\text{Detector}_1(c, x, n) := \text{true}\]
        \item \textbf{Temporal Nodes:} For $\mathsf{delay}$, it checks whether the new state (the current input) matches the stored state (the last input). Note that for level-2 delay, we technically need to compare two infinite streams. However, under IntConv, the stream of last input is always fixed after some point, so it can be stored in a finite state (see the example at the end of \secref{sec:intfp}).
        \[\text{Detector}_1(\mathsf{delay}, x, n) := (x(n) = z^{-1}(x)(n))\] 
        For $\mathsf{\lift{delay}}$, because its states are only updated within the inner time dimension, it has no states across the outer time dimension, so we always return true.
        \[\text{Detector}_1(\mathsf{\lift{delay}}, x, n) := \text{true}\]
        \item \textbf{Structural Combinators:} Call $\text{Detector}_1$ for its sub-circuits. That is 
        \[\text{Detector}_1(c_1 \seq c_2, x, n) := \text{Detector}_1(c_1, x, n) \land \text{Detector}_1(c_2, \means{c_1}(x), n)\] 
        and 
        \[\text{Detector}_1(c_1 \pll c_2, x, n) := \text{Detector}_1(c_1, x, n) \land \text{Detector}_1(c_2, x, n)\]
        \item \textbf{Feedback Loops:} For $\mathsf{loop}(c)$, it calls the detector on the loop body and checks whether the new state of the $z^{-1}$ on the back edge (the current output) matches the stored state (the delayed output). $o$ denotes the loop output $\means{\mathsf{loop}(c)}(x)$.
        \[
            \text{Detector}_1(\mathsf{loop}(c), x, n) :=
            \text{Detector}_1(c, (x, z^{-1}(o)), n) \land (o(n) = z^{-1}(o)(n))
        \] 
        For $\mathsf{\lift{loop}}(c)$, the $\lift{z^{-1}}$ on the back edge has no states across outer iterations, so it only calls the detector on the inner circuit. In the following, $o$ denotes $ \means{\mathsf{\lift{loop}}(c)}(x)$.
        \[\text{Detector}_1(\mathsf{\lift{loop}}(c), x, n) := \text{Detector}_1(c, (x, \lift{z^{-1}}(o)), n)\]
        \item \textbf{Level Adapters:} For $\mathsf{bracket}(c)$, it delegates to the internal circuit $c$: 
        \[\text{Detector}_1(\mathsf{bracket}(c), x, n) := \text{Detector}_1(c, \lift{\delta_0}(x), n)\]
        For $\mathsf{lifting}(c)$, it lacks states across outer iterations: 
        \[\text{Detector}_1(\mathsf{lifting}(c), x, n) := \text{true}\]
    \end{itemize}

    Similarly, we define $\text{Detector}_2$ as the level-2 StFP detector that accepts a level-2 circuit $c$, a nested stream input $x$, an index $(m, n)$, and returns a boolean. For standard and primitive nodes and structural combinators, it is defined by uniformly replacing $\text{Detector}_1$ with $\text{Detector}_2$ and the index $n$ with $(m, n)$. For other cases:
    \begin{itemize}
        \item \textbf{Temporal Nodes:} For $\mathsf{delay}$, it detects whether the delayed input from the last outer iteration has reached its fixpoint, which should have been stored in the state at the end of the last outer iteration (again, refer to the example at the end of \secref{sec:intfp}): 
        \[\text{Detector}_2(\mathsf{delay}, x, m, n) := \text{FixAfter}_2(z^{-1}(x), m, n)\] 
        For $\mathsf{\lift{delay}}$, the condition drops to just checking the state along the inner dimension: 
        \[\text{Detector}_2(\mathsf{\lift{delay}}, x, m, n) := (x(m, n) = \lift{z^{-1}}(x)(m, n))\]
        \item \textbf{Feedback Loops:} For $\mathsf{loop}(c)$, similar to $\mathsf{delay}$, it calls the detector on the internal circuit and checks whether the delayed output from the last outer iteration has reached its fixpoint. In the following, $o$ denotes $\means{\mathsf{loop}(c)}(x)$.
        \[\text{Detector}_2(\mathsf{loop}(c), x, m, n) := \text{Detector}_2(c, (x, z^{-1}(o)), m, n) \land \text{FixAfter}_2(z^{-1}(o), m, n)\] 
        For $\mathsf{\lift{loop}}(c)$, similar to $\mathsf{\lift{delay}}$, it calls the detector on the internal circuit, and checks whether the current output matches the delayed output. In the following, $o$ denotes $\means{\mathsf{\lift{loop}}(c)}(x)$.
        \[\text{Detector}_2(\mathsf{\lift{loop}}(c), x, m, n) := \text{Detector}_2(c, (x, \lift{z^{-1}}(o)), m, n) \land (o(m, n) = \lift{z^{-1}}(o)(m, n))\]
        \item \textbf{Level Adapters:} For $\mathsf{lifting}(c)$, it delegates to the level-1 detector: 
        \[\text{Detector}_2(\mathsf{lifting}(c), x, m, n) := \text{Detector}_1(c, x[m], n)\]
    \end{itemize}
\end{definition}

The $\text{FixAfter}_2$ conditions in the level-2 delay and loop clauses do not require predicting the future of the current outer iteration. Each is applied to a delayed stream, so its row at outer index $m$ comes from the already completed outer iteration $m-1$ (or is the zero row when $m=0$). Before iteration $m$ begins, the runtime has stored that row through a convergence bound $B$. If $n \geq B$, the bound immediately establishes $\text{FixAfter}_2$; otherwise, the runtime decides it by comparing the finitely many stored entries from $n$ through $B$, because all later entries equal the stored final value.

Like the StFP predicate, the StFP detector is structural and level-indexed. Its clauses for structural combinators recurse into sub-circuits, while its level-adapter clauses delegate across circuit levels. Therefore, it applies compositionally to nested bracketed circuits rather than treating each recursive component as an isolated case.

% An important fact related to implementation is that $\text{Detector}_2$ for the delay node needs to check whether the input from the last outer iteration $s$ is fixed after some index $n$. This is perfect for the sequential implementation because the computation of the last outer iteration has already terminated when we start computing the current outer iteration. For a parallel implementation, this is also totally acceptable, because we will know the result once the thread for the last iteration has terminated before $n$ or it has continued after $n$. Therefore, no computation would be stuck permanently. The example at the end of \secref{sec:intfp} will be helpful to understand this point.

We have proved the correctness of StFP via the following theorem:
\begin{theorem}[Correctness of StFP Detector]
    For any circuit $c$,
    \[\forall x,n,\, \text{Detector}_1(c, x, n) \iff \text{StFP}_1(c, x, n)\]
    and for any level-2 circuit $c$,
    \[\forall x,m,n,\, \text{Detector}_2(c, x, m, n) \iff \text{StFP}_2(c, x, m, n)\]
    These immediately imply that the StFP detector is both sound and complete.
\end{theorem}

The StFP detector should be read primarily as a correctness-oriented general fallback detector. It establishes that StFP is decidable for arbitrary circuits in our core calculus. A fixed-input check turns it into a local IntFP detector; applying that detector with zero-output checks across all bracketed sub-circuits yields the sound and complete global IntConv detector and establishes that IntConv is decidable as well. A practical implementation may optimize these checks or use a specialized detector when one is justified. In particular, for Conv-complete classes, IntConv and ExtConv coincide, so a sound and complete ExtConv detector also detects IntConv. FirstZero for the naive and semi-naive circuits is one such specialized detector. Measuring the overhead of these alternatives in a complete system is future work.

\section{IntConv is Complete for Useful Circuits}\label{sec:useful}
By Thm.~\ref{thm:intconv-sound}, IntConv is a sound sufficient condition for ExtConv. A natural question is: for what circuits is IntConv \emph{complete}? That is, when does mathematical convergence according to ExtConv also imply IntConv? For such circuits, a system following the IntConv criterion faithfully implements the theoretical specification. Moreover, because IntConv and ExtConv coincide on these circuits, a sound and complete ExtConv detector also detects IntConv. This permits specialized detectors such as FirstZero for the naive and semi-naive circuits instead of the general StFP detector from \secref{sec:detector}.

To answer this question, we introduce a key property called \emph{convergence-complete} (Conv-complete) circuits. They are just the circuits mentioned above for which IntConv is complete. In \secref{sec:regular-circuits}, we use \emph{regular circuits} as a name for the fragment generated by the standard DBSP constructs for lifted scalar computation, composition, Datalog recursion, and while recursion. The name lets us state a uniform completeness result for this existing recursive sublanguage: all regular circuits are Conv-complete (Thm.~\ref{thm:regular-convcomplete}).
Moreover, in \secref{sec:inc-opt}, we prove that the DBSP's incremental optimization ($\mathsf{plift}$ and $\mathsf{incOpt}$) preserves Conv-completeness (Thm.~\ref{thm:pushLifting-convcomplete} and Thm.~\ref{thm:incOpt-convcomplete}). As a result, all regular circuits and their incrementally optimized versions are Conv-complete.

Along with Conv-completeness, its local version FP-completeness is important for us to establish these results. We define them as follows:
\begin{definition}[Conv-complete Circuit]
A circuit $c$ is \emph{Conv-complete} iff for every input $x$,
\[
\text{ExtConv}(c, x) \implies \text{IntConv}(c, x).
\]
\end{definition}
\begin{definition}[FP-complete Circuit]
Let $c$ be a well-formed circuit.
We say $c$ is \emph{FP-complete} if:
\begin{itemize}
    \item \textbf{Level-1 completeness:}
    for every input $x$ and index $n$, if $\text{ExtConv}(c, x)$ and $\text{ExtFP}_1(c, x, n)$ hold, then there exists $n' \in \N$ such that $\text{IntFP}_1(c, x, n')$ holds.
    \item \textbf{Level-2 completeness (for level-2 circuits):}
    for every nested stream input $x$ and stream $\mathbf{b}$, if $\text{ExtConv}(c, x)$ and $\text{ExtFP}_2(c, x, \mathbf{b})$ hold, then there exists a stream $\mathbf{b}'$ such that $\text{IntFP}_2(c, x, \mathbf{b}')$ holds.
\end{itemize}
\end{definition}

We have established some compositional properties of Conv-completeness and FP-completeness. In particular, Conv-completeness is closed under all circuit constructs except for $\mathsf{bracket}$, while FP-completeness is closed under all circuit constructs except for $\mathsf{bracket}$ and sequential composition $\mathsf{seq}$. These results are useful for us to establish the important results in this section.

\subsection{Regular Circuits Are Conv-complete}\label{sec:regular-circuits}

Recall that we can use the DBSP circuit shown in Fig.~\ref{fig:naive-dbsp} to express a Datalog-style recursive query. We can express it in our formal language as follows.
\begin{definition}[Datalog Recursive Query]
    For a level-1 circuit $c_0$ that represents a lifted scalar query, the query body (the circuit between $\delta_0$ and $\elim$) of a Datalog-style recursive query is (the same as Eqn.~(\ref{eq:naive-body}))
    \[\mathsf{datalogBody}(c_0) := \Delta(\mathsf{loop}(c_0))\]
    and the streaming query (the lifted version of the whole circuit) is
    \[\mathsf{datalogQuery}(c_0) := \mathsf{bracket}(\mathsf{lifting}(\mathsf{datalogBody}(c_0)))\]
\end{definition}

Beyond recursive relational queries, the following \emph{while} program is used to show the Turing-completeness of DBSP~\cite{budiu2025dbsp}, where $Q$ is an arbitrary scalar function:
{\footnotesize
\begin{lstlisting}[language=Pascal]
x := i
while (x != Q(x))
  x := Q(x)
\end{lstlisting}
}
which can be implemented in DBSP as:
\begin{center}
\begin{tikzpicture}[>=latex]
    \node[] (input) {$i$};
    \node[block, right of=input] (delta) {$\delta_0$};
    \node[block, circle, right of=delta, inner sep=0cm] (p) {$+$};
    \node[block, right of=p] (Q) {$\lift Q$};
    \node[block, right of=Q] (D) {$\D$};
    \node[block, right of=D] (S) {$\elim$};
    \node[right of=S] (output)  {$x$};
    \node[block, below of=p, node distance=.7cm] (z) {$\zm$};
    \draw[->] (input) -- (delta);
    \draw[->] (delta) -- (p);
    \draw[->] (p) -- (Q);
    \draw[->] (Q) -- node (mid) {} (D);
    \draw[->] (D) -- (S);
    \draw[->] (mid.center) |- (z);
    \draw[->] (S) -- (output);
    \draw[->] (z) -- (p);
\end{tikzpicture}
\end{center}
In our formal language:
\begin{definition}[While Loop Programs]
    For a level-1 circuit $c_0$ representing a lifted scalar function, the query body of a while loop program is
    \[\mathsf{whileBody}(c_0) := \mathsf{loop}(\mathsf{add} \seq c_0) \seq \mathsf{D}\]
    and the streaming query is
    \[\mathsf{whileQuery}(c_0) := \mathsf{bracket}(\mathsf{lifting}(\mathsf{whileBody}(c_0)))\]
\end{definition}

Both query constructors above belong to the standard DBSP recursive sublanguage. Their results are level-1 circuits representing lifted scalar functions, so they can be composed or placed inside further recursive computations to form nested loops. We give the name \emph{regular circuits} to the sublanguage generated by these constructs so that we can state and prove FP- and Conv-completeness uniformly. We use \emph{lifted scalar circuit} for the base circuits that implement lifted scalar or relational queries.

\begin{definition}[Regular Circuit]
    \emph{Regular Circuits}, denoted by $\text{RegularCkt}$, is the class of level-1 circuits defined inductively as follows:
    \begin{itemize}
        \item \textbf{Lifted Scalar Circuit:} Any level-1 circuit composed (both sequentially and in parallel) of only primitive and standard nodes is a \emph{lifted scalar circuit}, which is also a regular circuit.
        \item \textbf{Sequence and Parallel Composition:} If $c_1$ and $c_2$ are regular circuits, their sequential composition $c_1 \seq c_2$ and parallel composition $c_1 \pll c_2$ are regular circuits.
        \item \textbf{Datalog Query:} Given a regular circuit $c_0$, $\mathsf{datalogQuery}(c_0)$ is a regular circuit.
        \item \textbf{While Loop:} Given a regular circuit $c_0$, $\mathsf{whileQuery}(c_0)$ is a regular circuit.
    \end{itemize}
\end{definition}

Using the inductive structure, it is easy to show that all regular circuits are FP-complete:
% Because all regular circuits represent a lifted scalar function, by leveraging Lemma~\ref{thm:lifted-scalar-FPcomplete},
%  it is easy to show that they are all FP-complete.
\begin{lemma}[Regular Circuits are FP-complete]
    Any regular circuit $c$ is FP-complete. In particular, for any $x, n$, $\text{ExtFP}_1(c, x, n) \implies \text{IntFP}_1(c, x, n)$.
\end{lemma}

We proceed to show that Datalog queries are Conv-complete. We first show that $\mathsf{datalogBody}$ satisfies a variant of FP-completeness tailored for a recursive query body, whose input must be in the form of $\delta_0(x)$ and the fixed output is zero.
\begin{lemma}[The iteration bound of $\mathsf{datalogBody}$]\label{thm:datalog-body-fpcomplete}
    If $c_0$ is a Conv-complete regular circuit, let $c := \mathsf{datalogBody}(c_0)$, for any scalar value $x$ and index $n$, we have
    \[ 
    \text{ExtConv}(c, \delta_0(x)) \land
    \text{ExtFP}_1(c, \delta_0(x), n) \land
    \means{c}(\delta_0(x))[n] = 0 \implies
    \text{IntFP}_1(c, \delta_0(x), n+1) \]
\end{lemma}

This lemma gives the concrete upper bound $n+1$ on the iteration at which IntFP holds.
It also allows us to prove $\mathsf{datalogQuery}$ is Conv-complete.
\begin{lemma}\label{thm:datalog-query-convcomplete}
    If $c_0$ is a Conv-complete regular circuit, then $\mathsf{datalogQuery}(c_0)$ is Conv-complete.
\end{lemma}

We have proved similar lemmas to establish the same iteration bound $n+1$ for $\mathsf{whileBody}$ and the Conv-completeness of $\mathsf{whileQuery}$.
These lemmas allow us to prove that regular circuits are Conv-complete by induction on the structure of regular circuits.
\begin{theorem}[Regular Circuits are Conv-complete]\label{thm:regular-convcomplete}
    Any regular circuit is Conv-complete.
\end{theorem}

From the proof above, we can see that the class of regular circuits is highly extensible. We can freely add more recursive circuit constructs similar to $\mathsf{datalogBody}$ and $\mathsf{whileBody}$ while preserving Conv-completeness, as long as we can establish the variant of FP-completeness in Thm.~\ref{thm:datalog-body-fpcomplete} for the new constructors.  

\subsection{Conv-complete Circuits Are Closed under Incremental Optimization}\label{sec:inc-opt}

As we can see in our example, the standard incremental optimization of DBSP can be represented by two syntactic transformations applied to a non-incremental circuit: $\mathsf{plift}$ to push the lifting operator onto individual nodes and $\mathsf{incOpt}$ to apply compositional incremental rules. In this section, we aim to show that $\mathsf{plift}$ and $\mathsf{incOpt}$ preserve Conv-completeness. To do this, we establish their correctness theorems to show how they transform the denotation, the IntFP, and the IntConv of the original circuit. The IntFP results track convergence bounds through these transformations; they are semantic guarantees rather than measurements of runtime cost.

To formally state the correctness of $\mathsf{plift}$, we introduce the notion of \emph{strong semantic equivalence} that requires the equivalence of both external denotation and internal convergence.
\begin{definition}[Strong Semantic Equivalence]
    For any circuit $c_1$ and $c_2$, we define the strong semantic equivalence $c_1 \simeq c_2$ as follows:
    \[\forall x,\, \means{c_1}(x) = \means{c_2}(x) \land (\text{IntConv}(c_1, x) \iff \text{IntConv}(c_2, x)) \]
    Note that this implicitly implies that $\text{ExtConv}(c_1, x) \iff \text{ExtConv}(c_2, x)$ because $\means{c}$ is partial.
\end{definition}

Based on this definition, we have re-established important equivalence lemmas in the DBSP theory, such as the chain rule $\Delta(c_1 \seq c_2) \simeq \Delta(c_1) \seq \Delta(c_2)$ and the lifting cycle rule $\mathsf{lifting}(\mathsf{loop}(c)) \simeq \mathsf{\lift{loop}}(\mathsf{lifting}(c))$. Using these, we prove the correctness of $\mathsf{plift}$ as follows. It is a straightforward induction on the structure of $c$.

\begin{theorem}[Correctness of $\mathsf{plift}$]\label{thm:pushLifting-correctness}
    For any level-1 circuit $c$,
    \begin{enumerate}
        \item (Denotation and IntConv) $\mathsf{lifting}(c) \simeq \mathsf{plift}(c)$.
        \item ($\text{IntFP}_1$) For any $x$ and $n$, $\text{ExtFP}_1(\mathsf{plift}(c), x, n) \implies \text{IntFP}_1(\mathsf{plift}(c), x, n)$.
        \item ($\text{IntFP}_2$) For any $x$ and $m,n \in \N$, $\text{IntFP}_1(c, x[m], n) \implies \text{IntFP}_2(\mathsf{plift}(c), x, m, n)$.
    \end{enumerate}
\end{theorem}

This allows us to prove that $\mathsf{plift}$ preserves both Conv-completeness and FP-completeness.
\begin{theorem}[$\mathsf{plift}$ preserves Conv-completeness and FP-completeness]\label{thm:pushLifting-convcomplete}
    For any level-1 circuit $c$, if $c$ is Conv-complete, then $\mathsf{plift}(c)$ is Conv-complete. If $c$ is FP-complete, then $\mathsf{plift}(c)$ is FP-complete.
\end{theorem}

For $\mathsf{incOpt}$, the situation is more tricky. In particular, unlike $\mathsf{lifting}(c)$ and $\mathsf{plift}(c)$, $\Delta(c)$ and $\mathsf{incOpt}(c)$ may have different IntConv behavior; just consider $\mathsf{node}(\mathsf{id})$ and $\Delta(\mathsf{node}(\mathsf{id}))$. This motivates us to introduce \emph{weak semantic equivalence}, which only requires the equivalence of external denotation but not internal convergence.
\begin{definition}[Weak Semantic Equivalence]
    For any circuit $c_1$ and $c_2$, we define the weak semantic equivalence $c_1 \approx c_2$ as $\forall x, \means{c_1}(x) = \means{c_2}(x)$. Note that this implies that $\text{ExtConv}(c_1, x) \iff \text{ExtConv}(c_2, x)$ since $\means{c}$ is a partial function.
\end{definition}

Besides denotation, how $\mathsf{incOpt}$ transforms IntFP is also tricky.
To see this, consider the simplest circuit $\mathsf{id}$, whose optimized incremental form is itself. Although the two forms are the same circuit, their corresponding input is different, because the input $x$ to the original form corresponds to the input $\D(x)$ to the incremental form. Because $x$ and $\D(x)$ have different FixAfter indices in general, the two forms have different IntFP in general.
This observation suggests that the formulation of the IntFP-preservation property that we desire for $\mathsf{incOpt}$ will be non-trivial.

An important observation is that although $x$ and $\D(x)$ may have different FixAfter indices, they do have a connection as follows:
\begin{proposition}
    For any stream $s$ and $n \in \N$, $\text{FixAfter}_1(s, n) \iff \text{ZeroAfter}_1(\D(s), n + 1)$,
    and for any nested stream $s$ and $\mathbf{b} \in \stream{\N}$, 
    $\text{FixAfter}_2(s, \mathbf{b}) \implies \text{FixAfter}_2(\D(s), \lift{\max}(\mathbf{b}, z^{-1}(\mathbf{b})))$.
\end{proposition}

Note that ZeroAfter implies FixAfter. This inspires us to raise a natural question: \emph{since input streams of $c$ and $\mathsf{incOpt}(c)$ have such FixAfter index relation, will all other intermediate streams share the same relation?} If so, as a result, IntConv should also be preserved. We formalize this idea as the following \emph{preservation} relation. 
\begin{definition}[IntFP and IntConv Preservation Relation]
    For any circuit $c_1$ and $c_2$, we say that $c_1$'s $\text{IntFP}_1$ is preserved by $c_2$, denoted by $c_1 \preserve_1 c_2$, if 
    \[\forall x, n,\, (\text{ExtConv}(c_1, x) \land \text{IntFP}_1(c_1, x, n)) \implies \text{IntFP}_1(c_2, \D(x), n + 1)\]
    For level-2 circuit $c_1$ and $c_2$, we say that $c_1$'s $\text{IntFP}_2$ is preserved by $c_2$, denoted by $c_1 \preserve_2 c_2$, if
    \[\forall x, \mathbf{b},\, (\text{ExtConv}(c_1, x) \land \text{IntFP}_2(c_1, x, \mathbf{b})) \implies \text{IntFP}_2(c_2, \D(x), \lift{\max}(\mathbf{b}, z^{-1}(\mathbf{b})))\]
    For any circuit $c_1$ and $c_2$, we say that $c_1$'s IntConv is preserved by $c_2$, denoted by $c_1 \preserve_{C} c_2$, if
    \[\forall x, \text{IntConv}(c_1, x) \implies \text{IntConv}(c_2, \D(x))\]
\end{definition}

These allow us to formally define what we expect from a ``good'' incremental optimization:
\begin{definition}[Sound Incremental Form]
    For a circuit $c_1$ and $c_2$, we say that $c_2$ is a \emph{sound incremental form} of $c_1$ if they satisfy the following four properties:
    \begin{enumerate}
        \item (Denotation) $\Delta(c_1) \approx c_2$.
        \item ($\text{IntFP}_1$) $c_1 \preserve_1 c_2$.
        \item ($\text{IntFP}_2$) $c_1 \preserve_2 c_2$ if $c_1$ is a level-2 circuit.
        \item (IntConv) $c_1 \preserve_{C} c_2$.
    \end{enumerate}
\end{definition}

Another important challenge is that $\mathsf{incOpt}$ is a flexible and extensible algorithm, as different primitive nodes can have different and custom incremental optimization. We deal with this via a uniform assumption that $\mathsf{incOpt}$ transforms any primitive node into a sound incremental form. This is a strong but still practical assumption, as we have established that the following common incremental forms are all sound:
\begin{proposition}[Common Sound Incremental Forms]\label{thm:common-sound-incremental-forms}
    \;
    \begin{enumerate}
        \item (Unoptimized) For any circuit $c$, $\Delta(c)$ is a sound incremental form of $c$.
        \item (Linear) For any linear function $f$ and $c = \mathsf{node}(f)$, $c$ is a sound incremental form of $c$.
        \item (Bilinear) For any bilinear function $f$ and $c = \mathsf{node}(f)$, the DBSP's bilinear incremental optimization~\cite{budiu2025dbsp} of $c$, denoted $\mathsf{bilinearOpt}(c)$, is a sound incremental form of $c$.
    \end{enumerate}
\end{proposition} 
The unoptimized incremental form provides a general sound fallback for any primitive node, ensuring the robustness of $\mathsf{incOpt}$. From now on, we may assume that $\mathsf{incOpt}$ transforms all its primitive nodes into a sound incremental form, and we can finally establish the correctness of $\mathsf{incOpt}$ as follows:
\begin{theorem}[Correctness of $\mathsf{incOpt}$]\label{thm:incOpt-correctness}
    For any circuit $c$, $\mathsf{incOpt}(c)$ is a sound incremental form of $c$.
\end{theorem}

This allows us to show that $\mathsf{incOpt}$ preserves Conv-completeness.
\begin{theorem}[$\mathsf{incOpt}$ preserves Conv-completeness]\label{thm:incOpt-convcomplete}
    For any circuit $c$, if $c$ is Conv-complete, then $\mathsf{incOpt}(c)$ is Conv-complete.
\end{theorem}

As a result, we can prove that the circuit implementing the delta-of-deltas strategy as shown in Fig.~\ref{fig:delta-of-deltas-dbsp} is Conv-complete.
\begin{proposition}
   If $c_0$ is a regular circuit, then $\mathsf{optDatalogQuery}(c_0)$ (Equation~\ref{eq:opt-datalog-query}) is Conv-complete.
\end{proposition}

\section{Lean Formalization}\label{sec:formalization}
All results in this paper have been formalized in Lean 4. We adapted the existing Lean 3 formalization of the DBSP theory~\cite{Chajed2022dbspF} to Lean 4 as a foundation, which accounts for 2,095 lines of code. On top of this, we formalized our new theory in 11,748 additional lines of code, covering the core IntConv theory, and its application such as a Hoare logic framework and several Zsets circuit case studies. These counts use \texttt{cloc} and exclude blank lines and comments.

The restriction of the stream level to $\ell \in \{1, 2\}$ in our formal language (\secref{sec:pre-language}) represents a deliberate trade-off between formalization effort (especially with mechanized proofs) and theoretical generality. This range is sufficient to express arbitrarily nested recursive queries and the delta-of-deltas incremental optimization, while keeping the formal verification tractable. In principle, there is no fundamental obstacle to generalizing $\ell$ to arbitrary natural numbers as our theory has already been developed in an inductive style (the level-indexed predicate).

\paragraph{Use of AI} AI is used to assist with Lean development and artifact preparation. The authors carefully reviewed all definitions and theorem statements and take full responsibility for the paper, formalization, and accompanying artifacts.
%See \secref{app:generalize-l} for more discussion.

\section{Related Work}\label{sec:related}
\paragraph{Incremental computation.}
Incremental computing is a paradigm for computing based on the idea of saving time by leveraging previous computational results.
It has a long history, going back to the 1964 and 1967 papers of Lombardi \cite{LISP:LR64,AIC:Lombardi67}.
There have been contributions from many fields, including theory
(dynamic graph algorithms \cite{ausiello1990dynamic,ramalingam1996incremental} and dynamic data structures \cite{bentley1980decomposable,leeuwen1980dynamization}), databases (view maintenance \cite{Book:GM99,chirkova2012materialized}), programming languages (incremental lambda-calculus \cite{field1990incremental}, incremental parsing \cite{ghezzi1979incremental}, interactive program-development environments \cite{reps1983incremental}, incremental dataflow analysis \cite{ryder1988incremental}, differential propagation in iterative fixpoint solvers \cite{fecht1998propagating}, and finite-differencing transformations in optimizing compilers \cite{paige1982finite}), document-preparation systems \cite{chen2000incremental}, constraint solvers \cite{freeman1990incremental}, truth-maintenance systems \cite{doyle1979truth}, and no doubt many others.
For a recent survey of incremental computation, see \cite{DBLP:conf/pepm/Liu24}.

Some of the work described above is only for limited models of computation.
DBSP is a Turing-complete language.
Because of its generality, DBSP could be applied to many of the application areas addressed by previous work, although it would not necessarily provide the same runtime guarantees of specialized results.
However, many of these applications would need the ability to detect fixpoints of nested computations.
The work described in this paper contributes to realizing the potential of DBSP by addressing an important semantic lacuna in the DBSP formalism.

More specifically, prior work has studied incremental recursive and fixpoint computation from complementary perspectives. Alvarez-Picallo et al.~\cite{AlvarezPicallo2019Fixing} develop a general theory of derivatives of fixpoints for incremental evaluation of recursive computations and Datalog. Abo Khamis et al.~\cite{AboKhamis2024Convergence} characterize algebraic conditions under which Datalog over pre-semirings converges in finitely many iterations and semi-naive evaluation applies. These works provide incremental fixpoint constructions or convergence characterizations for particular Datalog semantics. Our contribution is complementary: we formalize FPD as a finite-prefix runtime problem for incrementally transformed circuits and establish its general limitations and its exact detectability for useful circuit classes.

% There are several connections between incremental computing and memoization \cite{michie1968memo}, where previously computed results are systematically saved.
% Pugh and Teitelbaum \cite{pugh1989incremental} showed that by using non-standard representations of abstract data types (and choosing representation schemes that support unique representations), function caching provided a way to implement efficient incremental data-structure updates.
% Liu and Teitelbaum \cite{liu1995systematic} took these ideas a step further by identifying systematic methods to transform functional programs into incremental functional programs.
% A key step in their process was to discover sub-computations whose re-evaluation could be avoided by retrieving their values from cached results.
% Further developments along these lines were presented in Liu's book \cite{liu2013systematic}.
% \twr{We could probably eliminate this paragraph.}

% Persistent data structures \cite{driscoll1986making} provide an intriguing platform for incremental computing:
% with so-called \emph{full persistence}, one can both read and update any prior version of the computational state.
% This property is precisely what incremental computation needs: access to previous versions of the data without the cost of copying the whole structure.
% Okasaki \cite{Book:Okasaki99} showed that in a purely functional setting, persistence plus a careful treatment of lazy evaluation and memoization gives incremental computation for free.
% \twr{We could probably eliminate this paragraph.}

\paragraph{Feldera implementation of DBSP}
As mentioned in \secref{sec:intfp} and \secref{sec:detector}, the Feldera implementation of DBSP~\cite{feldera_repo} uses internal-state stability to detect convergence. Its runtime gives each operator a scope-indexed fixed-point check, intended to report whether the operator's output will remain stable under stable inputs, and declares a circuit stable when all of its operators report stability. Our theory gives this established implementation strategy a declarative account: IntFP and IntConv specify internal-state stability denotationally, StFP connects the declarative criterion to circuit state, and the StFP detector gives an algorithmic formalization.

The main novelty lies in the formal guarantees, especially the completeness result. We prove that IntConv is sound and that the StFP detector is sound and complete. The global IntConv detector combines the StFP detector with fixed-input and zero-output checks across all bracketed sub-circuits, yielding sound and complete IntConv detection. We also prove that IntConv is complete for a large class of useful circuits, including regular circuits and their incrementally optimized forms (\secref{sec:useful}). Thus, for these circuits, the internal-state-stability criterion detects every convergence admitted by the external specification. These results are machine-checked in Lean 4. Moreover, our level-indexed predicates and detectors give a formal, compositional account of nested bracketed circuits, whereas the Feldera SQL-to-DBSP compiler~\cite{feldera_repo} rejects a nested recursive operator within recursive code. Therefore, our theory may help guide future extended implementations.

\paragraph{Differential Dataflow and its mathematical foundation.}

Differential Dataflow (DD)~\cite{McSherry2013DifferentialD} also incrementalizes recursive queries over versions $T \times \N$, the DD counterpart of the delta-of-deltas setting in our counterexample. At the collection level, \texttt{Egress} selects the first repeated collection, at $i^* = \min\{i : W_{(t,i)} = W_{(t,i-1)}\}$; its differential implementation instead operates on $\delta W_{(t,i)}$. In this setting, $\delta W_{(t,i)}=0$ does not imply a repetition, since $W_{(t,i)}-W_{(t,i-1)}=\sum_{s\leq t}\delta W_{(s,i)}$. Thus, FirstZero cannot soundly determine $i^*$, and the paper defines differential \texttt{Egress} using $i^*$ without giving a finite-prefix criterion for obtaining it from the differential trace. Similar to DBSP, this also turns out to be a gap between the theory and the implementation of DD: its prototype instead reports convergence when no outstanding differences remain in the dataflow, a global internal-quiescence check similar in spirit to IntConv.

A later developed mathematical foundation for DD by Abadi et al.~\cite{Abadi2015FoundationsOD}, distinguishes fixed-iteration egress from first-repetition egress, but formalizes only the fixed-iteration policy; it notes that first-repetition egress would require partial streams. Consequently, its correctness theorem does not provide a compositional denotational account of data-dependent first-repetition termination for unbounded incremental iteration. Similarly, \textit{Flo}~\cite{Laddad2024FloAS}, a semantic foundation for progressive stream processing, requires bounded inner streams, leaving termination of unbounded inner iteration outside its scope. Our theory instead gives a sound criterion for such unbounded computations and identifies useful circuit classes for which that criterion is complete.

\section{Conclusion}\label{sec:conclusion}
In this paper, we developed a formal theory of convergence detection to address the fundamental gap between the mathematical specifications and practical implementations of Fixpoint Detection (FPD) in modern incremental recursive computation frameworks like DBSP. We identified and formalized the FPD problem, demonstrated that the commonly suggested ``FirstZero'' heuristic is unsound for the ``delta-of-deltas'' incremental optimization for recursive computation, and proved that exact FPD is impossible for arbitrary expressive circuits. Within our theory, we defined internal convergence (IntConv) as a declarative criterion corresponding to the internal-state-stability strategy used by practical implementations, and proved that it is a sound sufficient condition for convergence. We also defined the State Fixpoint (StFP) predicate and a sound and complete StFP detector, providing semantic and algorithmic formalizations of this strategy. The sound and complete global IntConv detector combines the StFP detector with fixed-input and zero-output checks across all bracketed sub-circuits. These predicates and detectors treat nested bracketed circuits compositionally. Furthermore, we proved that IntConv is complete for a broad class of useful circuits, including arbitrarily nested while loops, Datalog queries, and their incrementally optimized versions. These results provide formal semantic guarantees for an existing implementation strategy, and our results have been formalized in Lean 4.

Our results are mainly theoretical, rather than an empirical performance evaluation. A natural direction for future work is to implement and compare the global IntConv detector based on the general StFP detector with specialized global detectors within a complete system. Such an evaluation should measure detector overhead and compare end-to-end incremental execution with non-incremental recomputation, since that comparison also depends on the query, workload, data representation, and other system-level optimizations.

\section*{Data-Availability Statement}\label{sec:data-availability}
The Lean 4 formalization, including the impossibility result \Cref{thm:fpd-undecidability}, is available with build instructions at \url{https://github.com/Arcadia-Y/fixing-the-fixpoint/}. The repository also provides a paper-to-Lean mapping that links the paper's definitions and results to their formal declarations. The development is described in \secref{sec:formalization}.

\bibliographystyle{ACM-Reference-Format}
\bibliography{main}

%\appendix
%\input{appendix}

\end{document}